\documentstyle[aps,preprint]{revtex}
\def\f{\frac}

\begin{document}
{\sf
\title{
{\normalsize
\begin{flushright}
CU-TP972\\
RBRC-89
\end{flushright}}
A New Method to Derive Low-Lying N-dimensional Quantum Wave Functions 
by Quadratures  Along a Single Trajectory\thanks{Work supported in part by 
the U.S. Department of Energy}}

\author{
R. Friedberg$^{1}$, T. D. Lee$^{1-3}$ and W. Q. Zhao$^{2,4}$\\
{\small \it 1. Physics Department, Columbia University, New York, NY 10027, USA}\\
{\small \it 2. China Center of Advanced Science and Technology (CCAST)}\\
{\small \it         (World Lab.), P.O. Box 8730, Beijing 100080,  China}\\
{\small \it 3. RIKEN BNL Research Center (RBRC), Brookhaven National Lab.}\\
{\small \it         Bldg. 510, BNL, Upton, NY 11973, USA}\\
{\small \it 4. Institute of High Energy Physics, Academia Sinica, Beijing 
100039, China}  }
\maketitle

\newpage
\begin{abstract}
We present a new method to derive low-lying N-dimensional quantum wave 
functions by quadrature along a single trajectory. The N-dimensional
Schroedinger equation is cast into a series of readily integrable first
order ordinary differential equations. Our approach resembles the 
familiar W.K.B. approximation in one dimension, but is 
designed to explore the classically forbidden region 
and has a much wider applicability than W.K.B.. The method also provides a 
perturbation series expansion and the Green's functions of the wave
equation in N-dimension, all by quadratures along
a single trajectory. A number of examples are given for illustration, 
including a simple algorithm to evaluate the Stark effect in closed form 
to any finite order of the electric field.
\end{abstract}
\vspace{2cm}

~~~~PACS{:~~11.10.Ef,~~03.65.Ge}

\newpage

\section*{\bf 1. Introduction}
\setcounter{section}{1}
\setcounter{equation}{0}

In this paper, we begin with the generalization to arbitrary dimensions of a 
new method\cite{FLZ} in which the low-lying quantum wave function of a 
particle in 
a potential $+V$ can be obtained by quadratures along the trajectory
of a corresponding classical mechanical problem with $-V$ as its 
potential. Let ${\bf q}=(q_1, q_2, \cdots, q_N)$ describes the 
$N$-dimensional coordinates of the particle, and $\Phi({\bf q})$ is the 
ground-state (or low-lying) quantum wave function that satisfies
\begin{eqnarray}\label{e1.1}
H\Phi({\bf q}) = E\Phi({\bf q}),
\end{eqnarray}
where 
\begin{eqnarray}\label{e1.2}
H = -\frac{1}{2} {\bf \nabla}^2 + V({\bf q})
\end{eqnarray}
is the Hamiltonian of a non-relativistic particle of unit mass, and
\begin{eqnarray}\label{e1.3}
{\bf \nabla}^2 = \sum_{i=1}^{N}\frac{\partial^2}{\partial q_i^2}. 
\end{eqnarray}
We assume $V({\bf q})$ to be bounded from below; therefore we can choose 
\begin{eqnarray}\label{e1.4}
V({\bf q}) \geq 0.
\end{eqnarray}
Its minimum $V({\bf q}) = 0 $ may occur at more than one point, each with a 
non-vanishing curvature.

As in \cite{FLZ}, we introduce a scale factor $g^2$ by writing
\begin{eqnarray}\label{e1.5}
V({\bf q}) = g^2 v({\bf q}),
\end{eqnarray}
and consider the case of large $g$. We express 
\begin{eqnarray}\label{e1.6}
\Phi({\bf q}) = e^{-g {\bf S}({\bf q})}
\end{eqnarray}
in terms of a formal expansion:
\begin{eqnarray}\label{e1.7}
g {\bf S}({\bf q}) = g{\bf S}_0({\bf q}) + {\bf S}_1({\bf q}) + 
g^{-1}{\bf S}_2({\bf q}) + g^{-2}{\bf S}_3({\bf q}) + \cdots
\end{eqnarray}
with a similar expansion for $E$,
\begin{eqnarray}\label{e1.8}
E = gE_0 + E_1 + g^{-1}E_2 + g^{-2}E_3 + \cdots.
\end{eqnarray}
Substituting (\ref{e1.6}) - (\ref{e1.8}) into the Schroedinger equation 
(\ref{e1.1})
and equating the coefficients of $g^{-n}$ on both sides, we derive
\begin{eqnarray}\label{e1.9}
({\bf \nabla S}_0)^2 &=& 2v, \nonumber\\
{\bf \nabla S}_0 \cdot {\bf \nabla S}_1 &=& \frac{1}{2}{\bf \nabla}^2{\bf S}_0
-E_0,\nonumber\\
{\bf \nabla S}_0 \cdot {\bf \nabla S}_2 &=& \frac{1}{2} [{\bf \nabla}^2
{\bf S}_1 - ({\bf \nabla S}_1)^2] -E_1,\\
{\bf \nabla S}_0 \cdot {\bf \nabla S}_3 &=& \frac{1}{2} [{\bf \nabla}^2
{\bf S}_2 - 2({\bf \nabla S}_1) \cdot ({\bf \nabla S}_2)] -E_2, \nonumber
\end{eqnarray}
etc., where ${\bf \nabla}$ is the $N$-dimensional gradient vector whose 
components are $\nabla_i = \partial/\partial q_i$ and 
\begin{eqnarray}\label{e1.10}
i =1, 2, \cdots, N. 
\end{eqnarray}
The underlying reason for considering the expressions (\ref{e1.7}) and
(\ref{e1.8}) is that for large $g$ the ground-state quantum wave function
has its amplitude centered around the minimum of $V({\bf q})$; near the 
minimum, $V({\bf q})$ may be approximated by a quadratic function of 
${\bf q}$, with the curvature of the corresponding potential surface 
proportional to $g^2$. Thus, for large $g$, the ground-state energy is 
$O(g)$ and so is the exponent of the ground-state wave function, as
suggested  by the ground-state energy and the corresponding wave function
of a harmonic oscillator. The 
parameter $g^{-1}$ serves as a measure of anharmonicity.

It is useful to write the first line of (\ref{e1.9}) as
\begin{eqnarray}\label{e1.11}
\frac{1}{2}({\bf \nabla S}_0)^2 - v({\bf q}) = e,
\end{eqnarray}
where 
\begin{eqnarray}\label{e1.12}
e = 0+.
\end{eqnarray}
Equation (\ref{e1.11}) is equivalent to  the corresponding 
Hamilton-Jacobi equation in classical mechanics, with $-V({\bf q})$ 
as the potential and $e$ the total
energy. To solve $(\ref{e1.11})$, we pick a minimum of $V({\bf q})$, say 
${\bf q} = {\bf 0}$ with
\begin{eqnarray}\label{e1.13}
V({\bf 0}) = g^2 v({\bf 0}) = 0.
\end{eqnarray}
Consider a trajectory ${\bf q}(t)$ which begins at ${\bf q}={\bf 0}$ 
when $t=0$ and ends at ${\bf q}_T$ when $t=T$; i.e.,
\begin{eqnarray}\label{e1.14}
{\bf q}(0) = {\bf 0}~~~~~~~~{\sf and}~~~~~~~~~~{\bf q}(T)={\bf q_T}.
\end{eqnarray}

Hamilton's action-integral is given by the minimum of the path integral
\begin{eqnarray}\label{e1.15}
{\bf S}_0({\bf q_T}, e) = \int\limits_0^T [\frac{1}{2}\sum_{i=1}^N 
\stackrel{\bf {\cdot}}{q}_i(t)^2 - (-v({\bf q}(t)))]dt + T~e,
\end{eqnarray}
where
\begin{eqnarray}\label{e1.16}
T = (\frac{\partial {\bf S}_0}{\partial e})_{{\bf q_T}}
\end{eqnarray}
is the total time of the trajectory from the initial point {\bf 0} 
to the final 
point ${\bf q_T}$. Since the potential in the classical problem 
$= -v({\bf q}) \leq 0$, the integrand in (\ref{e1.15}) is positive 
everywhere. Hence, given any final point ${\bf q_T}$ and any total 
energy $e > 0$, the minimum of (\ref{e1.15}) exists, so is therefore 
the classical trajectory. (For clarity, assume the minimum to be unique,
a restriction that will be removed later.) As $e \rightarrow 0+$, 
the initial velocity $\stackrel{{\bf \cdot}}{{\bf q}}$ is 
proportional to $e^{1/2}$ which also approaches zero; the total time $T$ 
of the classical trajectory is $\propto e^{-1/2}\rightarrow \infty$, 
yielding 
\begin{eqnarray}\label{e1.17}
\lim_{e=0+} T~e = 0.
\end{eqnarray}
The same limit $e=0+$ of (\ref{e1.15}) gives ${\bf S}_0$ of 
(\ref{e1.11}) - (\ref{e1.12}).

To go back to the first line of (\ref{e1.9}), we designate the end-point 
${\bf q_T}$ of the classical trajectory simply as ${\bf q}$, the point 
of interest for the quantum solution $\Phi({\bf q})$. Setting 
\begin{eqnarray}\label{e1.18}
{\bf q_T} = {\bf q},
\end{eqnarray}
the solution ${\bf S}_0({\bf q})$ of the first line in (\ref{e1.9}) can now be 
written as
\begin{eqnarray}\label{e1.19}
{\bf S}_0({\bf q}) = \int\limits_0^T [\frac{1}{2}\sum_{i=1}^N 
\stackrel{\bf {\cdot}}{q}_i^2 - (-v({\bf q}))]dt,
\end{eqnarray}
where the integral is along the trajectory $q_i(t)$ that satisfies the 
classical equations of motion
\begin{eqnarray}\label{e1.20}
\stackrel{{\bf \cdot \cdot}}q_i = \partial v/\partial q_i,
\end{eqnarray}
the energy conservation
\begin{eqnarray}\label{e1.21}
\frac{1}{2}\sum_{i=1}^N \stackrel{\bf {\cdot}}{q}_i(t)^2 -v({\bf q}(t)) = 0+
\end{eqnarray}
and the initial and final configurations (\ref{e1.14}) and (\ref{e1.18}).

The solution ${\bf S}_0({\bf q})$ thus 
obtained is positive and continuous everywhere; it is an increasing function 
along the direction of the classical trajectory. In the case that the 
classical potential $-v({\bf q})$ has other absolute maxima besides $0$, 
say
\begin{eqnarray}\label{e1.22}
-v({\bf a_1}) = -v({\bf a_2}) = \cdots = -v({\bf a_n}) =  0,
\end{eqnarray}
${\bf S}_0({\bf q})$ is not analytic at these isolated points
\begin{eqnarray}\label{e1.23}
{\bf q} = {\bf a_1}, {\bf a_2}, \cdots, {\bf a_n}.
\end{eqnarray}
For a total energy $e > 0$, along the classical trajectory the velocity 
must continue in its direction when the trajectory passes through any of 
the other maxima of $-v({\bf q})$, say ${\bf q} = {\bf a_n}$. Thus, as 
$e \rightarrow 0+$, ${\bf \nabla}{\bf S}_0$ becomes zero at 
${\bf q} = {\bf a_n}$, but its sign along the trajectory-direction
remains the {\it same} before or after 
${\bf a_n}$; this forces ${\bf \nabla}{\bf S}_0$ to develop a kink at 
${\bf q} = {\bf a_n}$. At the initial point ${\bf q} = {\bf 0}$,
while $-v({\bf 0})=0$ and therefore ${\bf \nabla S}_0=0$ as $e=0+$,  
${\bf S}_0({\bf q})$ is analytic, since different trajectories 
emanating from 
${\bf q} = {\bf 0}$ have to go along different directions. At infinity, 
it is easy to see that ${\bf S}_0({\bf q}) = \infty$, and therefore 
$\Phi_0 \equiv e^{-g{\bf S}_0}$ is zero.

To solve the second equation in (\ref{e1.9}), we require ${\bf S}_1({\bf q})$
to be also analytic at ${\bf q} = {\bf 0}$; this yields
\begin{eqnarray}\label{e1.24}
E_0 = \frac{1}{2}({\bf \nabla}^2{\bf S}_0)_{{\rm at}~{\bf q}={\bf 0}}.
\end{eqnarray}
It is convenient to consider the surface
\begin{eqnarray}\label{e1.25}
{\bf S}_0({\bf q}) = {\sf constant};
\end{eqnarray}
its normal is along the corresponding classical trajectory passing through 
${\bf q}$. Characterize  each classical trajectory by the ${\bf S}_0$-value
along the trajectory and a set of $N-1$ angular variables
\begin{eqnarray}\label{e1.26}
\alpha = (\alpha_1({\bf q}),\alpha_2({\bf q}), \cdots, \alpha_{N-1}({\bf q})),
\end{eqnarray}
so that each $\alpha$ determines one classical trajectory with 
\begin{eqnarray}\label{e1.27}
{\bf \nabla} \alpha_j \cdot {\bf \nabla} {\bf S}_0 = 0,
\end{eqnarray}
where
\begin{eqnarray}\label{e1.28}
j = 1, 2, \cdots, N-1.
\end{eqnarray}
(As an example, we note that  as ${\bf q} \rightarrow {\bf 0}$, 
$v({\bf q}) \rightarrow \frac{1}{2} \sum\limits_i \omega_i^2 q_i^2$
and therefore 
$ {\bf S}_0 \rightarrow \frac{1}{2} \sum\limits_i \omega_i q_i^2$.
Consider  the ellipsoidal surface 
$ {\bf S}_0 =$ constant $\equiv \frac{1}{2} l^2$. For $l$ sufficiently small,
each classical trajectory is normal to this ellipsoidal surface.
A convenient choice of $\alpha$ could be simply any $N-1$ orthogonal 
parametric coordinates on the surface.) 
Each $\alpha$ designates one classical trajectory, and vice versa. 
Every $({\bf S}_0, \alpha)$ is mapped into a unique set 
$(q_1,~q_2,~\cdots,~q_N)$ with ${\bf S}_0 \geq 0$
by construction. Depending on the problem, the converse may or may not 
be true; i.e. $(q_1,~q_2,~\cdots,~q_N) \rightarrow ({\bf S}_0, \alpha)$ 
could be either one to one, or one to many. In the latter case, we regard 
the points in the ${\bf q}$-space as specified by the coordinates 
$({\bf S}_0, \alpha)$, rather than the original coordinates
$(q_1,~q_2,~\cdots,~q_N)$.

Write
\begin{eqnarray}\label{e1.29}
{\bf S}_1({\bf q}) = {\bf S}_1({\bf S}_0, \alpha), 
\end{eqnarray}
the second line of (\ref{e1.9}) becomes
\begin{eqnarray}\label{e1.30}
({\bf \nabla S}_0)^2 (\frac{\partial {\bf S}_1}{\partial {\bf S}_0})_{\alpha} 
= \frac{1}{2}{\bf \nabla}^2 {\bf S}_0 -E_0,
\end{eqnarray}
and therefore
\begin{eqnarray}\label{e1.31}
{\bf S}_1({\bf q}) = {\bf S}_1({\bf S}_0, \alpha) = \int\limits_0^{{\bf S}_0}
\frac{d{\bf S}_0}{({\bf \nabla S}_0)^2} [\frac{1}{2} {\bf \nabla}^2{\bf S}_0 -
E_0],
\end{eqnarray}
where the integration is taken along the classical trajectory of constant
$\alpha$. Likewise, the third, fourth and other lines of (\ref{e1.9}) 
lead to
\begin{eqnarray}\label{e1.32}
E_1 &=& \frac{1}{2} [{\bf \nabla}^2{\bf S}_1 - ({\bf \nabla S}_1)^2 ]_
{{\rm at}~{\bf q}={\bf 0}},\\
{\bf S}_2({\bf q}) = {\bf S}_2({\bf S}_0, \alpha) &=& \int\limits_0^{{\bf S}_0}
\frac{d{\bf S}_0}{({\bf \nabla S}_0)^2} \{\frac{1}{2} [{\bf \nabla}^2
{\bf S}_1 - ({\bf \nabla S}_1)^2] - E_1\},\\
E_2 &=& \frac{1}{2} [{\bf \nabla}^2{\bf S}_2 - 2 ({\bf \nabla S}_1)\cdot
({\bf \nabla S}_2) ]_{{\rm at}~{\bf q}={\bf 0}},\\
{\bf S}_3({\bf q}) = {\bf S}_3({\bf S}_0, \alpha) &=& \int\limits_0^{{\bf S}_0}
\frac{d{\bf S}_0}{({\bf \nabla S}_0)^2} \{\frac{1}{2} [{\bf \nabla}^2
{\bf S}_2 - 2 ({\bf \nabla S}_1) \cdot ({\bf \nabla S}_2)] - E_2\},
\end{eqnarray}
etc. These solutions give the convenient normalization convention at 
${\bf q} = {\bf 0}$,
\begin{eqnarray}\label{e1.36}
&{\bf S}({\bf 0}) = 0 \nonumber\\
{\rm and}~~~~~~~~~~~~~~~~~~~~~~\\
&\Phi({\bf 0}) = 1.\nonumber
\end{eqnarray}

The above expressions for $E_0,~E_1,~E_2,~\cdots$ can be derived more 
directly by introducing
\begin{eqnarray}\label{e1.37}
{\bf \sigma} ({\bf q}) &\equiv& g {\bf S}({\bf q}) - g~{\bf S}_0({\bf q})
\nonumber\\
     &=& {\bf S}_1({\bf q}) +g^{-1}{\bf S}_2({\bf q}) +
     g^{-2}{\bf S}_3({\bf q}) + \cdots .
\end{eqnarray}
Thus, the ground-state wave function $\Phi({\bf q})$ is
\begin{eqnarray}\label{e1.38}
\Phi = e^{-g{\bf S}_0-{\bf \sigma}},
\end{eqnarray}
and the corresponding Schroedinger equation (\ref{e1.1}) gives
\begin{eqnarray}\label{e1.39}
g{\bf \nabla}{\bf S}_0 \cdot {\bf \nabla} {\bf \sigma} = \frac{1}{2}
[g{\bf \nabla}^2 {\bf S}_0
+{\bf \nabla}^2 {\bf \sigma} -({\bf \nabla} {\bf \sigma})^2] - E.
\end{eqnarray}
At ${\bf q}={\bf 0}$, $v({\bf q})=0$ and ${\bf \nabla S}_0=0$, on account of 
$({\bf \nabla S}_0)^2=2v$. Assuming ${\bf \nabla} \sigma$ to be regular, 
we have
\begin{eqnarray}\label{e1.40}
E = \frac{1}{2}[g{\bf \nabla}^2 {\bf S}_0
+{\bf \nabla}^2 {\bf \sigma} -({\bf \nabla} {\bf \sigma})^2]_
{{\sf at}~{\bf q}={\bf 0}}
\end{eqnarray}
which leads to (\ref{e1.24}), (\ref{e1.32}), (1.34), $\cdots$ for 
$E_0$, $E_1$, $E_2$, $\cdots$.

\noindent
\underline{Example}.

As an example, consider the special case
\begin{eqnarray}\label{e1.41}
V({\bf q}) = \frac{g^2}{2}(q_1^2+q_2^2+\cdots +q_N^2).
\end{eqnarray}
From (\ref{e1.9}), one can readily show that the first line gives
\begin{eqnarray}\label{e1.42}
{\bf S}_0({\bf q}) = \frac{g}{2}(q_1^2+q_2^2+\cdots +q_N^2)
\end{eqnarray}
and (\ref{e1.24}) leads to 
\begin{eqnarray}\label{e1.43}
E_0 = N/2;
\end{eqnarray}
the other equations (\ref{e1.31}) - (1.35), etc., yield 
$E_1=E_2=\cdots=0$ and $S_1=S_2=\cdots=0$. The result is the well known 
exact answer
\begin{eqnarray}\label{e1.44}
\Phi({\bf q}) = exp[-\frac{g}{2}(q_1^2+q_2^2+\cdots +q_N^2)]
\end{eqnarray}
and
\begin{eqnarray}\label{e1.45}
E = \frac{g}{2}N.
\end{eqnarray}

The organization of the paper is as  follows: In Section 2, we discuss 
the case of  a separable potential, and show that the solution of the
wave function is a product of  one-dimensional functions, as expected.
The particular one-dimensional example of
\begin{eqnarray}\label{e1.46}
V(x) = \frac{g^2}{2}(x^2-a^2)^2
\end{eqnarray}
has been analyzed in ref.\cite{FLZ} using our method.
In this problem, besides the 
anharmonicity parameter $g^{-1}$, there is  also a much smaller barrier 
penetration parameter
\begin{eqnarray}\label{e1.47}
e^{-\frac{4}{3}ga^3}.
\end{eqnarray}
(The same problem has also been extensively studied in the literature[2-9], 
using mostly Feynman's path integration method. Quite often, the two 
parameters $g$ and $a$ in (\ref{e1.46}) are related by $2ag=1$, which makes 
$V=\frac{1}{2}y^2(1-gy)^2$ with $y=x+a$. The dimensionless anharmonicity
parameter $1/ga^3$ then becomes $8g^2$ and the barrier penetration 
parameter (\ref{e1.47}) becomes $e^{-1/6g^2}$. In our 
$g^{-1}$-expansion, we keep the parameter $a$ fixed.)

The ground-state wave  function $\psi_{even}(x)$ is even in $x$, and
the first excited state is an odd function $\psi_{odd}(x)$. Let 
$E_{even}$ and $E_{odd}$ be their corresponding eigenvalues. We can
expand both their sum and difference as the following  double  series:
\begin{eqnarray}\label{e1.48}
\frac{1}{2}(E_{odd}+E_{even}) =
ga\sum\limits_{m,n} C_{mn}(ga^3)^{-m} e^{-\frac{4}{3}nga^3} 
\end{eqnarray}
and
\begin{eqnarray}\label{e1.49}
\frac{1}{2}(E_{odd}-E_{even}) &=&
4(\frac{2}{\pi}g^3a^5)^{\frac{1}{2}} e^{-\frac{4}{3}ga^3}\nonumber\\
&&\cdot 
\sum\limits_{m,n} c_{mn}(ga^3)^{-m} e^{-\frac{4}{3}(n-1)ga^3} 
\end{eqnarray}
where $C_{mn}$ and $c_{mn}$  are numerical coefficients, which can be 
explicitly expressed as definite integrals\cite{FLZ}, with the results
\begin{eqnarray}\label{e1.50}
c_{m0} &=& 0~~~~~~~~~~{\sf for~all}~~m,\nonumber\\
C_{00} &=&  c_{01} = 1,~~~~~~C_{10} = -\frac{1}{4},\\
C_{20} &=& -\frac{9}{2^6},~~~~~~~~~C_{30} = -\frac{89}{2^9},
~~~{\sf etc.}\nonumber
\end{eqnarray}
The  terms  with $n=1$ represent the one-instanton contributions[2-9],
$n=2$ the two-instanton contributions, etc. As one can  
see  from the  derivation,  an important tool, in addition  to the
$g^{-1}$ expansion (\ref{e1.7}) - (\ref{e1.9}) discussed  above,  is
the use  of the Sturm-Liouville type  of  Green's functions. In  this
paper, we  will  show how to  generalize  such Green's function  technique
to an N-dimensional problem. In Section 3, we introduce a compact notation
which will pave the way towards the generalization.

Section 4 contains the main body of this paper. We depart from the initial 
approach based on the Hamilton-Jacobi equation; instead, we consider
two N-dimensional Hamiltonians:
\begin{eqnarray}\label{e1.51}
H=-\frac{1}{2}{\bf \nabla}^2 + V({\bf q})
\end{eqnarray}
with, as before,
\begin{eqnarray}
V({\bf q}) = g^2 v({\bf q}) \geq  0,\nonumber
\end{eqnarray}
and
\begin{eqnarray}\label{e1.52}
{\cal H}=  H + \epsilon U({\bf q}).
\end{eqnarray}
Assuming that the ground-state wave function 
$\Phi({\bf q}) = e^{-g {\bf S}({\bf q})}$ of the unperturbed Hamiltonian 
$H$ is known, we set out to derive the perturbation series expansion  
for the corresponding  ground-state wave function $\Psi({\bf q})$ of
${\cal H}$ with $\epsilon U$ as the perturbation. Again, as we shall see, 
to each  order  in $\epsilon$, the  perturbed $\Psi({\bf q})$ can be
obtained by quadrature along a  single trajectory connecting ${\bf 0}$,
the point that the unperturbed $\Phi({\bf q})$ is maximum (i.e.  
${\bf S}$ is minimum), to the point ${\bf q}$ of interest; the trajectory 
is determined by being the normal to the ${\bf S}$ = constant surfaces 
along the entire course of  the  trajectory. (It may be called the
quantum trajectory, in contrast to the classical trajectory determined  
by the Hamilton-Jacobi equation.) The form of the new perturbation  
series differs from  the usual expression; the result is of course the
same, as will be illustrated by a few examples. The  new perturbation 
series formula will also be shown to cover the previous analysis in
Sections 1-3 based on
the  classical trajectory derived  from  the  Hamilton-Jacobi 
equation  as special cases.

In the same Section 4, the Green's function $G$ of the N-dimensional  
Hamiltonian $H$  will  be  derived, with $G$ satisfying
\begin{eqnarray}\label{e1.53}
(H-E)G={\sf I}
\end{eqnarray}
and $E$ being the ground-state energy of $H$. Usually, the Green's 
function and the perturbation series expansion to each order 
$\epsilon^n$, require either an infinite sum over all excited levels  
of $H$,  or equivalently, a sum over all possible paths through  
Feynman's path integration method. The Green's function and the
perturbation series formula derived here are different; they depend only
on quadratures along a  single trajectory.

In Section 5, we give examples to illustrate the applications of the new
perturbation series, as well as to clarify the properties of the Green's
function. In Section 6, we show how the trajectory-quadrature formulation 
can be extended to derive excited states, and in Section 7 we discuss how 
generalization can be 
made to allow for singular potentials, such as an
attractive Coulomb potential with perturbations. As we shall see, with 
our new approach we can evaluate , e.g., the Stark effect to any finite 
order of the perturbation by quadratures along the radial trajectory.


Although our methods are elementary, they become surprisingly powerful 
for the following reason. For example, in (1.38) by writing the wave 
function in the form of $e^{-g{\bf S}_0-\sigma}$, where $g$ is large, we 
cause the cross-term $g{\bf \nabla}{\bf S}_0 \cdot {\bf \nabla}\sigma$
to be much larger than 
$\frac{1}{2}[{\bf \nabla}^2\sigma - ({\bf \nabla}\sigma)^2]$.
Therefore the differential equation for $\sigma$ can be written order by 
order in such a way that 
$\frac{1}{2}[{\bf \nabla}^2\sigma - ({\bf \nabla}\sigma)^2]$
is known from previous orders, and  
${\bf \nabla}{\bf S}_0 \cdot {\bf \nabla}\sigma = ({\bf \nabla}{\bf S}_0)^2
\frac{\partial \sigma}{\partial {\bf S}_0}$ is inferred from the equation. 
In effect, to each order, $\sigma$ now satisfies a first order ordinary 
differential equation (along the trajectory of constant $\alpha$), not a 
second order partial differential equation. This makes possible a direct
solution through quadratures. With variations, this theme runs through all 
the sections to be presented in this paper (sometimes $e^{-\sigma}$ is 
written $e^{-\tau}$, or $\chi$, and sometimes the known exponent is called
$g{\bf S}$, rather than $g{\bf S}_0$).


\section*{\bf {2. Separable Potential}}
\setcounter{section}{2}
\setcounter{equation}{0}

For clarity, we consider the special case of dimension $N=2$. The discussions
below for a separable potential can be trivially extended to any $N>2$.
Let the two dimensional coordinate ${\bf q}$ in the previous section
be designated
\begin{eqnarray}\label{e2.1}
{\bf q} = (x, y)
\end{eqnarray}
and the quantum mechanical potential $V({\bf q})$ and ${\bf \nabla}^2$ in
(\ref{e1.2}) be
\begin{eqnarray}\label{e2.2}
V({\bf q}) = g^2 v({\bf q}) = g^2 [v_x(x) + v_y(y)]
\end{eqnarray}
and
\begin{eqnarray}\label{e2.3}
{\bf \nabla}^2 = (\partial^2/\partial x^2) + (\partial^2/\partial y^2).
\end{eqnarray}
As in (\ref{e1.4}) and (\ref{e1.13}), we assume
\begin{eqnarray}\label{e2.4}
v_x(x) \geq 0,~~~~~~~~ v_y(y) \geq 0
\end{eqnarray}
and at the point $x=y=0$,
\begin{eqnarray}\label{e2.5}
v_x(0) = v_y(0) = 0.
\end{eqnarray}

The classical Lagrangian density for the corresponding potential
$-v({\bf q})$ is
\begin{eqnarray}\label{e2.6}
\frac{1}{2} [\stackrel{{\bf \cdot}}{x}(t)^2+\stackrel{{\bf \cdot}}{y}(t)^2]
+ v_x(x) + v_y(y),
\end{eqnarray}
where the dots denote time-derivatives, as before. The equations of motion 
are
\begin{eqnarray}\label{e2.7}
&\stackrel{{\bf \cdot \cdot}}{x}(t) = \frac{d}{dx}v_x(x)\nonumber\\
{\rm and}~~~~~~~~~~~~~~~~~~~~~ &\\
&\stackrel{{\bf \cdot \cdot}}{y}(t) = \frac{d}{dy}v_y(y).\nonumber
\end{eqnarray}
As in (\ref{e1.14}), we consider a trajectory that begins at
$x(0) = y(0) = 0$ (the maximum, or one
of the maxima, of $-v$) at $t=0$ and ends at
\begin{eqnarray}\label{e2.8}
x(T) = x_T~~~{\rm and}~~~y(T) = y_T
\end{eqnarray}
when $t=T>0$, in which, $x_T$ and $y_T$ are variables independent of T. The
action integral along the trajectory is given by the sum of
\begin{eqnarray}\label{e2.9}
&A_x(x_T,T) = \int\limits_0^T [\frac{1}{2} \stackrel{{\bf \cdot}}{x}^2 -
(-v_x(x))]dt \nonumber\\
{\rm and}~~~~~~~~~~~~~~~~~~~~~ &\\
&A_y(y_T,T) = \int\limits_0^T [\frac{1}{2} \stackrel{{\bf \cdot}}{y}^2 -
(-v_y(y))]dt \nonumber
\end{eqnarray}
The minimum of $A_x(x_T, T)$ and $A_y(y_T, T)$, keeping $x_T$, $y_T$ and $T$
fixed, gives the Lagrangian equations of motion (\ref{e2.7}), and the
derivatives of $A_x$ and $A_y$ with respect to the final variables
$x_T$, $y_T$ and $T$ are
\begin{eqnarray}\label{e2.10}
&\frac{\partial}{\partial x_T} A_x(x_T,T) = \stackrel{{\bf \cdot}}{x}(T)
\nonumber\\
&\frac{\partial}{\partial y_T} A_y(y_T,T) = \stackrel{{\bf \cdot}}{y}(T)
\nonumber\\
&\frac{\partial}{\partial T} A_x(x_T,T) = -e_x \\
{\rm and}~~~~~~~~~~~~~~~~~~& \nonumber \\
&\frac{\partial}{\partial T} A_y(y_T,T) = -e_y, \nonumber
\end{eqnarray}
where $e_x$ and $e_y$ are the energies in the $x$ and $y$ directions, given by
\begin{eqnarray}\label{e2.11}
&e_x = \frac{1}{2} \stackrel{{\bf \cdot}}{x}^2 - v_x(x)
\nonumber\\
{\rm and}~~~~~~~~~~~~~~~~~~& \\
&e_y = \frac{1}{2} \stackrel{{\bf \cdot}}{y}^2 - v_y(y).
\nonumber
\end{eqnarray}
For a given final
configuration characterized by $x_T$, $y_T$ and $T$, both $e_x$, $e_y$,
and therefore also their sum
\begin{eqnarray}\label{e2.12}
e = e_x + e_y
\end{eqnarray}
are determined. Following (\ref{e1.15}), we transform the independent
variables from $x_T$, $y_T$ and $T$ to $x_T$, $y_T$ and $e$. We write
\begin{eqnarray}\label{e2.13}
S_0 (x_T, y_T, e) = A_x(x_T,T) + A_y(y_T,T) + T~e,
\end{eqnarray}
where $T$ is a dependent variable, determined by
\begin{eqnarray}\label{e2.14}
T = T(x_T,y_T,e) = (\frac{\partial S_0}{\partial e})_{x_T,y_T}.
\end{eqnarray}
Given $x_T$, $y_T$ and $e$, the classical trajectory is given by
the minimum
of (\ref{e2.13}), from which $T$, $e_x$, $e_y$ are also determined. Thus,
we can regard
\begin{eqnarray}\label{e2.15}
&e_x = e_x(x_T,y_T,e), \nonumber \\
{\rm and}~~~~~~~~~~~~~~~~& \\
&e_y = e_y(x_T,y_T,e).\nonumber
\end{eqnarray}
Hamilton's action integral (\ref{e1.15}) becomes
\begin{eqnarray}\label{e2.16}
S_0(x_T,y_T,e) = X_0(x_T,e_x) + Y_0(y_T,e_y),
\end{eqnarray}
where 
\begin{eqnarray}\label{e2.17}
&X_0(x_T,e_x) = A_x(x_T,T) + T~e_x \nonumber \\
{\rm and}~~~~~~~~~~~~~~~~&  \\
&Y_0(y_T,e_y) = A_y(y_T,T) + T~e_y, \nonumber 
\end{eqnarray}
in which $T$ is given by (\ref{e2.14}).
Combining (\ref{e2.10}) and (\ref{e2.17}), we can
readily show that
\begin{eqnarray}\label{e2.18}
&(\frac{\partial S_0}{\partial x_T})_{y_T,e} =
(\frac{\partial X_0}{\partial x_T})_{e_x} =
(\frac{\partial A_x}{\partial x_T})_T = \stackrel{{\bf \cdot}}{x}(T)
\nonumber \\
&(\frac{\partial S_0}{\partial y_T})_{x_T,e} =
(\frac{\partial Y_0}{\partial y_T})_{e_y} =
(\frac{\partial A_y}{\partial y_T})_T = \stackrel{{\bf \cdot}}{y}(T)\\
{\rm and}~~~~~~~~~~~~~~~~& \nonumber \\
&(\frac{\partial X_0}{\partial e_x})_{x_T} =
(\frac{\partial Y_0}{\partial e_y})_{y_T} =
(\frac{\partial S_0}{\partial e})_{x_T,y_T} = T. \nonumber
\end{eqnarray}

Along each classical trajectory, when the final time variable changes
from $T$ to $T+dT$, the end point moves from $(x_T,y_T)$ to
($x_T+\stackrel{{\bf \cdot}}{x}(T)dT$, $y_T+\stackrel{{\bf \cdot}}{y}(T)dT$);
correspondingly,
\begin{eqnarray}\label{e2.19}
dS_0 = (\frac{\partial X_0}{\partial x_T})_{e_x}
\stackrel{{\bf \cdot}}{x}(T)dT +
(\frac{\partial Y_0}{\partial y_T})_{e_y} \stackrel{{\bf \cdot}}{y}(T)dT
+e~dT
\end{eqnarray}
Combining the first two lines of (\ref{e2.18}) with (\ref{e2.19}) and
taking the limit $e=0+$, we find
\begin{eqnarray}\label{e2.20}
dS_0 &=& [(\frac{\partial X_0}{\partial x_T})^2 +
(\frac{\partial Y_0}{\partial y_T})^2]dT \nonumber \\
&=& ({\bf \nabla} S_0)^2 dT.
\end{eqnarray}
Along the trajectory we also have
\begin{eqnarray}\label{e2.21}
dT = dx_T/\stackrel{{\bf \cdot}}{x}(T) = dy_T/\stackrel{{\bf \cdot}}{y}(T),
\end{eqnarray}
which, together with (\ref{e2.18}), leads to
\begin{eqnarray}\label{e2.22}
dT = dx_T/ (\frac{\partial X_0}{\partial x_T})_{e_x} =
dy_T/ (\frac{\partial Y_0}{\partial y_T})_{e_y}.
\end{eqnarray}
Substituting (\ref{e2.20}) and (\ref{e2.22}) into (\ref{e1.31}) and 
labeling $x_T$, $y_T$ simply as $x$, $y$, we obtain
\begin{eqnarray}\label{e2.23}
S_1(x,y) = X_1(x) + Y_1(y),
\end{eqnarray}
where
\begin{eqnarray}\label{e2.24}
X_1(x) = \int\limits_0^x dx (\frac{\partial X_0}{\partial x})^{-1}
[\frac{1}{2}\frac{\partial^2 X_0}{\partial x^2} -E_{0,x}],\nonumber \\
Y_1(x) = \int\limits_0^y dy (\frac{\partial Y_0}{\partial y})^{-1}
[\frac{1}{2}\frac{\partial^2 Y_0}{\partial y^2} -E_{0,y}]
\end{eqnarray}
with
\begin{eqnarray}\label{e2.25}
&E_{0,x} = \frac{1}{2}
(\frac{\partial^2 X_0}{\partial x^2})_{{\rm at}~x=0}\nonumber \\
{\rm and}~~~~~~~~~~~~~~~~~~~~& \\
&E_{0,y} = \frac{1}{2}
(\frac{\partial^2 Y_0}{\partial y^2})_{{\rm at}~y=0}\nonumber 
\end{eqnarray}
and
\begin{eqnarray}\label{e2.26}
E_{0} = E_{0,x} + E_{0,y},
\end{eqnarray}
consistent with (\ref{e1.24}).

Likewise, ${\bf S}_2$, ${\bf S}_3$, $\cdots$ can be written as
sums of the form $X_2(x)+Y_2(y)$, $X_3(x)+Y_3(y)$, $\cdots$, with
\begin{eqnarray}\label{e2.27}
X_2(x) &=& \int\limits_0^x dx (\frac{\partial X_0}{\partial x})^{-1}
\{\frac{1}{2}[\frac{\partial^2 X_1}{\partial x^2} -
(\frac{\partial X_1}{\partial x})^2] - E_{1,x}\},\nonumber\\
Y_2(y) &=& \int\limits_0^y dy (\frac{\partial Y_0}{\partial y})^{-1}
\{\frac{1}{2}[\frac{\partial^2 Y_1}{\partial y^2} -
(\frac{\partial Y_1}{\partial y})^2] - E_{1,y}\},\\
X_3(x) &=& \int\limits_0^x dx (\frac{\partial X_0}{\partial x})^{-1}
\{\frac{1}{2}[\frac{\partial^2 X_2}{\partial x^2} -
2\frac{\partial X_1}{\partial x}\frac{\partial X_2}{\partial x}] - E_{2,x}\},
\nonumber\\
Y_3(y) &=& \int\limits_0^y dy (\frac{\partial Y_0}{\partial y})^{-1}
\{\frac{1}{2}[\frac{\partial^2 Y_2}{\partial y^2} -
2\frac{\partial Y_1}{\partial y}\frac{\partial Y_2}{\partial y}] - E_{2,y}\},
\nonumber
\end{eqnarray}
where
\begin{eqnarray}\label{e2.28}
E_{1,x} &=& \frac{1}{2}[\frac{\partial^2 X_1}{\partial x^2} -
(\frac{\partial X_1}{\partial x})^2]_{{\sf at}~x=0},\nonumber\\
E_{1,y} &=& \frac{1}{2}[\frac{\partial^2 Y_1}{\partial y^2} -
(\frac{\partial Y_1}{\partial y})^2]_{{\sf at}~y=0},\\
E_{2,x} &=& \frac{1}{2}[\frac{\partial^2 X_2}{\partial x^2} -
2\frac{\partial X_1}{\partial x}
\frac{\partial X_2}{\partial x}]_{{\sf at}~x=0},\nonumber\\
E_{2,y} &=& \frac{1}{2}[\frac{\partial^2 Y_2}{\partial y^2} -
2\frac{\partial Y_1}{\partial y}
\frac{\partial Y_2}{\partial y}]_{{\sf at}~y=0},\nonumber
\end{eqnarray}
etc.. The corresponding wave function factors $e^{-g{\bf S}_0}$,
$e^{-{\bf S}_1}$, $e^{-g^{-1}{\bf S}_2}$, $\cdots$ are all products
$e^{-g(X_0+Y_0)}$, $e^{-(X_1+Y_1)}$, $e^{-g^{-1}(X_2+Y_2)}$, etc.. 


\section*{\bf {3. A Compact Notation}}
\setcounter{section}{3}
\setcounter{equation}{0}

In this section we shall go over completely to the variables
${\bf S}_0$, $\alpha$ and express the $g^{-1}$ expansion 
((\ref{e1.7}) and (\ref{e1.31})-(1.35)) described in Section 1 as
iteration of an integral operator along a classical trajectory, that is
for variable ${\bf S}_0$ and fixed $\alpha$. As we shall see, our results 
can be written in a compact form by introducing an implicit matrix
notation in which the indices are two values of  ${\bf S}$.

In terms of the function ${\bf \sigma}({\bf q})$, introduced by
(\ref{e1.37}), the wave function $\Phi({\bf q})$ is
\begin{eqnarray}\label{e3.1}
\Phi({\bf q}) = e^{-g{\bf S}_0({\bf q})-{\bf \sigma}({\bf q})}.
\end{eqnarray}
Because 
\begin{eqnarray}\label{e3.2}
({\bf \nabla S}_0)^2 &=& 2v, 
\end{eqnarray}
the Schroedinger equation (\ref{e1.1}) is equivalent to the following
equation for $e^{-{\bf \sigma}}$:
\begin{eqnarray}\label{e3.3}
[g({\bf \nabla}{\bf S}_0) \cdot {\bf \nabla}+T]e^{- {\bf \sigma}({\bf q})} =
[-\frac{g}{2}{\bf \nabla}^2 {\bf S}_0 + E]e^{- {\bf \sigma}({\bf q})},
\end{eqnarray}
where
\begin{eqnarray}\label{e3.4}
T=-\frac{1}{2} {\bf \nabla}^2.
\end{eqnarray}
Its solution is given by (\ref{e1.37}) and (\ref{e1.40}), which can be
written in a more compact form, as we shall see.

Consider the coordinate transformation
\begin{eqnarray}\label{e3.5}
q_1, q_2, q_3, \cdots, q_N \rightarrow
{\bf S}_0, \alpha_1, \alpha_2, \cdots, \alpha_{N-1}
\end{eqnarray}
with $\alpha=(\alpha_1, \alpha_2, \cdots, \alpha_{N-1})$ denoting the set
of $N-1$ orthogonal angular coordinates introduced in
(\ref{e1.26})~-~(\ref{e1.28}). Define the $\theta_0$-function:
\begin{eqnarray}\label{e3.6}
\theta_0({\bf S}_0-{\bf {\overline S}}_0)=
\left\{\begin{array}{cc}
1 & ~~~~~~~~{\sf if}\hspace{4mm} 0 \leq {\bf {\overline S}}_0 < {\bf S}_0 \\
0 & ~~~~~~~~{\sf if} \hspace{4mm} 0 \leq {\bf S}_0 < {\bf {\overline S}}_0  
\end{array}
\right. 
\end{eqnarray}
where ${\bf S}_0$ and ${\bf {\overline S}}_0$ vary from $0$ to $\infty$.
For ${\bf S}_0 > 0$, its derivative gives Dirac's $\delta$-function:
\begin{eqnarray}\label{e3.7}
\frac{\partial}{\partial {\bf S}_0}
\theta_0({\bf S}_0-{\bf {\overline S}}_0) =
\delta({\bf S}_0-{\bf {\overline S}}_0).
\end{eqnarray}
Let us define $C_0(\alpha)$ as a square matrix in the ${\bf S}_0$-space, 
with its matrix element
\begin{eqnarray}\label{e3.8}
({\bf S}_0|C_0(\alpha)|{\bf {\overline S}}_0) =
g^{-1}\theta_0({\bf S}_0 -
{\bf {\overline S}}_0)/({\bf \nabla}{\bf {\overline S}}_0)^2.
\end{eqnarray}
An important feature is that each matrix element of $C_0(\alpha)$ connects
two points ${\bf q}({\bf S}_0, \alpha)$ and
${\overline {\bf q}}({\overline {\bf S}}_0, \alpha)$ along the same
classical trajectory that satisfies (\ref{e1.20})~-~(\ref{e1.21}), and
therefore having the same angular variables
$\alpha_1, \alpha_2, \cdots, \alpha_{N-1}$. Throughout the paper, the
variables ${\bf S}_0$ and ${\overline {\bf S}}_0$ vary between $0$ and
$\infty$, with  the end points $0$ and $\infty$ treated as boundaries.
All gradients refer to differentiations with respect to the Cartesian
coordinates; i.e.,
\begin{eqnarray}\label{e3.9}
{\bf \nabla}_j = \frac{\partial}{\partial q_j}~~~~~~{\sf and}~~~~~~
{\bf \nabla}^2 = \sum_{j=1}^{N}\frac{\partial^2}{\partial q_j^2}, 
\end{eqnarray}
as before. On account of (\ref{e1.27}) and (\ref{e3.6}),
for ${\bf S}_0 >0$,
\begin{eqnarray}\label{e3.10}
{\bf \nabla}{\bf S}_0 \cdot 
{\bf \nabla}\theta_0({\bf S}_0-{\bf {\overline S}}_0)=
({\bf \nabla}{\bf S}_0)^2 \delta({\bf S}_0-{\bf {\overline S}}_0),
\end{eqnarray}
where the $j^{{\sf th}}$ component of 
${\bf \nabla}\theta_0({\bf S}_0-{\bf {\overline S}}_0)$ is
$${\bf \nabla}_j\theta_0({\bf S}_0-{\bf {\overline S}}_0)=
\frac{\partial}{\partial q_j}\theta_0({\bf S}_0-{\bf {\overline S}}_0)$$
with $q_j=q_j({\bf S}_0, \alpha)$. From (\ref{e3.8}), we also have
\begin{eqnarray}\label{e3.11}
g{\bf \nabla}{\bf S}_0 \cdot
{\bf \nabla}({\bf S}_0|C_0(\alpha)|{\bf {\overline S}}_0)=
\delta({\bf S}_0-{\bf {\overline S}}_0).
\end{eqnarray}

Like $C_0(\alpha)$, $\theta_0$ also denotes a square matrix whose matrix
elements are
\begin{eqnarray}\label{e3.12}
({\bf S}_0|\theta_0|{\bf {\overline S}}_0)=
\theta_0({\bf S}_0-{\bf {\overline S}}_0).
\end{eqnarray}
Matrix multiplications follow the usual rule; e.g.,
\begin{eqnarray}\label{e3.13}
({\bf S}_0|C_0 \theta_0|{\bf {\overline S}}_0)=
\int_0^{\infty}d{\overline {\overline {\bf S}}}_0
({\bf S}_0|C_0|{\bf {\overline {\overline S}}}_0)
( {\overline {\overline {\bf S}}}_0|\theta_0|{\bf {\overline S}}_0).
\end{eqnarray}
Introduce $h_0^2(\alpha)$ to be a diagonal matrix whose elements are
\begin{eqnarray}\label{e3.14}
({\bf S}_0|h_0^2(\alpha)|{\bf {\overline S}}_0)=
\delta({\bf S}_0-{\overline {\bf S}}_0)/
({\bf \nabla} {\bf {\overline S}}_0 )^2.
\end{eqnarray}
Thus, 
\begin{eqnarray}\label{e3.15}
({\bf S}_0|\theta_0 h_0^2(\alpha)|{\bf {\overline S}}_0)=
\theta_0({\bf S}_0-{\overline {\bf S}}_0)/
({\bf \nabla} {\overline {\bf S}}_0 )^2,
\end{eqnarray}
but
\begin{eqnarray}\label{e3.16}
({\bf S}_0|h_0^2(\alpha)\theta_0|{\bf {\overline S}}_0)=
\theta_0({\bf S}_0-{\overline {\bf S}}_0)/
({\bf \nabla} {\bf S}_0 )^2.
\end{eqnarray}
Equations (\ref{e3.8}) and(\ref{e3.11}) can be written in their matrix
forms
\begin{eqnarray}\label{e3.17}
C_0 = g^{-1} \theta_0 h_0^2
\end{eqnarray}
and 
\begin{eqnarray}\label{e3.18}
g{\bf \nabla} {\bf S}_0 \cdot {\bf \nabla} C_0 = 1,
\end{eqnarray}
in which the $\alpha$-dependences of $C_0$ and $h_0^2$ are suppressed.

\noindent
\underline{Theorem 1.} The solution of the wave equation
(\ref{e3.3})~-~(\ref{e3.4}) for $e^{-{\bf \sigma}}$ satisfies
\begin{eqnarray}\label{e3.19}
e^{-{\bf \sigma}({\bf S}_0,~\alpha)}&=& 1+ \int_0^{\infty}
d{\bf {\overline S}}_0
({\bf S}_0|[1+C_0(\alpha) T]^{-1}C_0(\alpha)|{\bf {\overline S}}_0)
\nonumber\\
&&~~~~[-\frac{g}{2} {\bf \nabla}^2 {\bf {\overline S}}_0 +E]
e^{-{\bf \sigma}({\overline {\bf S}}_0,~\alpha)}.
\end{eqnarray}
\underline{Proof} (i) We first establish the existence of the inverse
matrix $[1+C_0(\alpha) T]^{-1}$. Assuming the opposite, there would be a
column matrix $f(\alpha)$ which satisfies
\begin{eqnarray}\label{e3.20}
[1+C_0(\alpha) T] f(\alpha) = 0.
\end{eqnarray}
Applying ${\bf \nabla} {\bf S}_0 \cdot {\bf \nabla}$ from the left, using
(\ref{e3.18}) and on account of $T=-\frac{1}{2} {\bf \nabla}^2$, we have
for the matrix element $f({\bf S}_0,~\alpha)$ of $f(\alpha)$:
\begin{eqnarray}\label{e3.21}
g{\bf \nabla} {\bf S}_0 \cdot {\bf \nabla} f({\bf S}_0,~\alpha)
-\frac{1}{2}{\bf \nabla}^2 f({\bf S}_0,~\alpha) = 0.
\end{eqnarray}
Hence,
\begin{eqnarray}\label{e3.22}
{\bf \nabla}\cdot [e^{-2g{\bf S}_0} {\bf \nabla} f({\bf S}_0,~\alpha)] =0.
\end{eqnarray}
Multiplying this equation by $f({\bf S}_0, \alpha)$ and integrating over
all space, we derive, through partial integration,
\begin{eqnarray}\label{e3.23}
\int d^N {\bf q}~e^{-2g{\bf S}_0} [{\bf \nabla} f({\bf S}_0,~\alpha)]^2=0,
\end{eqnarray}
provided
\begin{eqnarray}\label{e3.24}
e^{-2g{\bf S}_0} f {\bf \nabla} f = 0~~~~~~~{\sf at}~~~~~~~~~
{\bf S}_0 = \infty.
\end{eqnarray}
It is clear that (\ref{e3.23}) implies 
${\bf \nabla} f({\bf S}_0,~\alpha)=0$; i.e.,
$f({\bf S}_0,~\alpha)=$ constant, which makes
$-\frac{1}{2}{\bf \nabla}^2 f = 0$. It follows then from (\ref{e3.20}),
\begin{eqnarray}
f({\bf S}_0,~\alpha) =0.\nonumber
\end{eqnarray}
(ii) From (\ref{e3.3}),
\begin{eqnarray}\label{e3.25}
[g({\bf \nabla}{\bf S}_0) \cdot {\bf \nabla}+T](e^{- {\bf \sigma}}-1) =
[-\frac{g}{2}{\bf \nabla}^2 {\bf S}_0 + E]e^{- {\bf \sigma}}.
\end{eqnarray}
Next, applying $g({\bf \nabla}{\bf S}_0) \cdot {\bf \nabla}$ onto
$1+C_0T$ and using (\ref{e3.18}), we find
\begin{eqnarray}\label{e3.26}
(g({\bf \nabla}{\bf S}_0) \cdot {\bf \nabla})[1+C_0T]=
g({\bf \nabla}{\bf S}_0) \cdot {\bf \nabla} +T;
\end{eqnarray}
therefore, (\ref{e3.25}) becomes
\begin{eqnarray}\label{e3.27}
[g({\bf \nabla}{\bf S}_0) \cdot {\bf \nabla}][1+C_0T](e^{- {\bf \sigma}}-1) =
[-\frac{g}{2}{\bf \nabla}^2 {\bf S}_0 + E]e^{- {\bf \sigma}}.
\end{eqnarray}
On the other hand, assuming (\ref{e3.19}) we have
\begin{eqnarray}\label{e3.28}
e^{- {\bf \sigma}}-1 = (1+C_0T)^{-1} C_0
[-\frac{g}{2}{\bf \nabla}^2 {\bf S}_0 + E]e^{- {\bf \sigma}}
\end{eqnarray}
which satisfies (\ref{e3.27}). In deriving the above, we see that for 
$f=e^{-\sigma} -1$, (\ref{e3.24}) is satisfied, which ensures the 
applicability of $ (1+C_0T)^{-1}$. The theorem is then proved.

According to (\ref{e3.17}), $C_0=O(g^{-1})$, and therefore, for $g^2>>1$,
we may expand
\begin{eqnarray}\label{e3.29}
(1+C_0T)^{-1} C_0=C_0+(-C_0T)C_0+(-C_0T)^2C_0+ \cdots.
\end{eqnarray}
Neglecting $O(g^{-1})$, $\sigma \cong {\bf S}_1$ and approximating
$(1+C_0T)^{-1}C_0$ by $C_0$, (\ref{e3.28}) reduces to
\begin{eqnarray}\label{e3.30}
e^{- {\bf S}_1}-1 =
C_0[-\frac{g}{2}{\bf \nabla}^2 {\bf S}_0 + E]e^{- {\bf S}_1}.
\end{eqnarray}
Taking its derivative through $g{\bf \nabla}{\bf S}_0 \cdot {\bf \nabla}$,
we have, as expected,
\begin{eqnarray}\label{e3.31}
g[{\bf \nabla}{\bf S}_0 \cdot {\bf \nabla}{\bf S}_1]e^{- {\bf S}_1} =
[\frac{g}{2}{\bf \nabla}^2 {\bf S}_0 - E]e^{- {\bf S}_1}
\end{eqnarray}
which, upon the approximation $E \cong gE_0$, gives the second line of
(\ref{e1.9}). Likewise, we can derive (\ref{e1.31})~-~(1.35) for
${\bf S}_1$, $E_1$, ${\bf S}_2$, $E_2$, 
$\cdots$ from (\ref{e3.28}).

\newpage

\section*{\bf {4. Green's Functions and Perturbation Series}}
\setcounter{section}{4}
\setcounter{equation}{0}

As we have seen, the $g^{-1}$ expansion enables us to probe the
ground-state wave function $\Phi({\bf q})$ along a classical trajectory;
to each order $g^{-n}$, the $N$-dimensional quantum wave function can
be calculated through quadratures, by performing a finite number of
definite integrals along the same trajectory. To the same order of accuracy,
the corresponding energy is also determined by a finite number of
differentiations at a single point  where the potential has an absolute 
minimum, say at ${\bf q} = {\bf 0}$. This result may seem unfamiliar, 
since the
Schroedinger wave function $\Phi({\bf q})$ and its eigenvalue $E$ depend
sensitively on the boundary conditions at $\infty$. But in our approach,
the determination of $\Phi({\bf q})$ is based only on a single trajectory
connecting ${\bf 0}$ and ${\bf q}$. The boundary condition at infinity
appears almost automatically, because the classical action (i.e., Hamilton's
action integral) satisfies the property:
\begin{eqnarray}\label{e4.1}
{\bf S}_0({\bf q}) \rightarrow \infty~~~~~~~~~{\sf as}~~~~~
{\bf q}\rightarrow \infty.
\end{eqnarray}
Since the classical action is calculated for a potential $-V({\bf q})$,
which has its maximum at $0$, this condition (\ref{e4.1}) imposes only a
general restriction on the potential; in any case, it is considered to be
an input in the new method. To explore further the essential features of the 
underlying
mechanism, we shall in this section  generalize our formalism 
{\it without} any explicit use of the Hamilton-Jacobi equation.

In this section we shall derive several new forms of the Green's function
for the N-dimensional Schroedinger equation; these are all based on
quadratures along a single specified trajectory. With the Green's function
we will be able to arrive at a new perturbation series expansion, different
from the usual familiar ones. As we shall see (near the end of the section),
the material discussed in previous sections 1-3 will appear as special cases.
As mentioned in the Introduction, the new Green's function will also enable
us to extend the multi-instanton calculations developed for the
one-dimensional problem in ref.\cite{FLZ} 
to arbitrary dimensions. In what follows, we shall first give a number of
definitions, then for clarity, arrange the main results in the form of
three theorems (Theorem 2, 3 and 4 below). Examples of the new 
perturbation series and the Green's
function will be given in Section 5.

Consider two $N$-dimensional Hamiltonians:
\begin{eqnarray}\label{e4.2}
H = -\frac{1}{2} {\bf \nabla}^2 + V({\bf q})
\end{eqnarray}
with $V({\bf q}) \geq 0$, as before, and
\begin{eqnarray}\label{e4.3}
{\cal H} = H +\epsilon U({\bf q}).
\end{eqnarray}
Their ground-state wave functions are $\Phi({\bf q})$ and $\Psi({\bf q})$
respectively:
\begin{eqnarray}\label{e4.4}
H\Phi({\bf q}) = E\Phi({\bf q})
\end{eqnarray}
and
\begin{eqnarray}\label{e4.5}
{\cal H}\Psi({\bf q}) = {\cal E} \Psi({\bf q}).
\end{eqnarray}
Let 
\begin{eqnarray}\label{e4.6}
\epsilon \Delta \equiv {\cal E} -E.
\end{eqnarray}
Write,  as before,
\begin{eqnarray}\label{e4.7}
V({\bf q}) = g^2 v({\bf q}),
\end{eqnarray}
\begin{eqnarray}\label{e4.8}
\Phi({\bf q}) = e^{-g{\bf S}({\bf q})}
\end{eqnarray}
and
\begin{eqnarray}\label{e4.9}
\Psi({\bf q}) = e^{-g{\bf S}({\bf q})-\tau ({\bf q})}.
\end{eqnarray}
We have, from (\ref{e4.4})~-~(\ref{e4.9}),
\begin{eqnarray}\label{e4.10}
\frac{g^2}{2}({\bf \nabla}{\bf S})^2 -
\frac{g}{2}{\bf \nabla}^2{\bf S} -g^2 v + E = 0
\end{eqnarray}
and
\begin{eqnarray}\label{e4.11}
g {\bf \nabla}{\bf S}\cdot {\bf \nabla}\tau +
\frac{1}{2}[({\bf \nabla}\tau)^2 - {\bf \nabla}^2 \tau] =
\epsilon (U - \Delta).
\end{eqnarray}
In the subsequent discussions in {\it this} section, we assume that
${\bf S}({\bf q})$ is known, and our objective is to compute
$\tau({\bf q})$, assuming that
\begin{eqnarray}\label{e4.12}
|\epsilon| << 1.
\end{eqnarray}

We assume further that
${\bf S}({\bf q})$ has an overall behavior similar to the
Hamilton-Jacobi action ${\bf S}_0({\bf q})$ given by
(\ref{e1.19}); in particular, ${\bf S}({\bf q})$ has an absolute minimum
at ${\bf q}={\bf 0}$. (This convention will be adopted here on.) Therefore
\begin{eqnarray}\label{e4.13}
{\bf \nabla}{\bf S} = {\bf 0}~~~~~~~~~~{\sf at}~~~~~~~~{\bf q}={\bf 0}.
\end{eqnarray}
(Note that from here on the absolute minimum of  $V({\bf q})$ may or
may  not be at the  same point.)
Throughout this section, the role of the Hamilton-Jacobi action integral
${\bf S}_0$ in previous sections will be replaces by ${\bf S}$.
As in (\ref{e1.26})~-~(\ref{e1.28}), we introduce a set of $N-1$ angular
variables
\begin{eqnarray}\label{e4.14}
\beta = (\beta_1({\bf q}),\beta_2({\bf q}), \cdots, \beta_{N-1}({\bf q})),
\end{eqnarray}
which satisfy
\begin{eqnarray}\label{e4.15}
{\bf \nabla} \beta_j \cdot {\bf \nabla} {\bf S} = 0
\end{eqnarray}
with
\begin{eqnarray}\label{e4.16}
j = 1, 2, \cdots, N-1.
\end{eqnarray}
Each point ${\bf q}$ in the $N$-dimensional space will now be designated by
\begin{eqnarray}\label{e4.17}
({\bf S}, \beta_1, \beta_2, \cdots, \beta_{N-1}),
\end{eqnarray}
instead of $(q_1, q_2, q_3, \cdots, q_N)$.
At ${\bf q}$, let $\stackrel{\wedge}{\bf S},~\stackrel{\wedge}{\beta}_1,
\stackrel{\wedge}{\beta}_2,\cdots,
\stackrel{\wedge}{\beta}_{N-1}$
form a set of $N$ orthonormal unit vectors, which are the normals of the
${\bf S}=$ constant, $\beta_1=$ constant, $\cdots$, $\beta_{N-1}=$ constant
surfaces at that point. Correspondingly, a line element 
$d\stackrel{\rightarrow}{{\bf q}}$ can be written as
\begin{eqnarray}\label{e4.18}
d\stackrel{\rightarrow}{{\bf q}} = 
\stackrel{\wedge}{\bf S} h_{\bf S} d{\bf S}
+\sum_{j=1}^{N-1}\stackrel{\wedge}{\beta}_j h_j d\beta_j;
\end{eqnarray}
the gradient is given by
\begin{eqnarray}\label{e4.19}
{\bf \nabla} = \stackrel{\wedge}{\bf S} 
\frac{1}{h_{\bf S}}\frac{\partial}{\partial {\bf S}}
+\sum_{j=1}^{N-1}\stackrel{\wedge}{\beta}_j \frac{1}{h_j}\frac{\partial}{\partial \beta_j},
\end{eqnarray}
and 
\begin{eqnarray}\label{e4.20}
T=-\frac{1}{2} {\bf \nabla}^2
\end{eqnarray}
can be decomposed into two parts:
\begin{eqnarray}\label{e4.21}
T = T_{\bf S} + T_{\beta},
\end{eqnarray}
with
\begin{eqnarray}\label{e4.22}
T_{\bf S} = -\frac{1}{2h_{\bf S}h_\beta} \frac{\partial}{\partial {\bf S}}
(\frac{h_{\beta}}{h_{\bf S}}\frac{\partial}{\partial {\bf S}}),
\end{eqnarray}
in which
\begin{eqnarray}\label{e4.23}
h_{\beta} = \prod_{j=1}^{N-1}h_j,
\end{eqnarray}
and
\begin{eqnarray}\label{e4.24}
T_{\beta} = -\frac{1}{2h_{\bf S}h_\beta} \sum_{j=1}^{N-1}
\frac{\partial}{\partial \beta_j}
(\frac{h_{\bf S}h_{\beta}}{h_j^2}\frac{\partial}{\partial \beta_j}).
\end{eqnarray}
It follows from (\ref{e4.18}),
\begin{eqnarray}\label{e4.25}
h_{\bf S}^2 = [({\bf \nabla}{\bf S})^2]^{-1},
h_1^2 = [({\bf \nabla}\beta_1)^2]^{-1},\cdots,
h_j^2 = [({\bf \nabla}\beta_j)^2]^{-1},\cdots.
\end{eqnarray}
The  volume element in the ${\bf q}$-space is
\begin{eqnarray}\label{e4.26}
d^N {\bf q} = h_{\bf S}h_\beta d{\bf S}d\beta
\end{eqnarray}
where
\begin{eqnarray}\label{e4.27}
d\beta = \prod_{j=1}^{N-1}d\beta_j.
\end{eqnarray}

The trajectory of a given $\beta$ defines a continuous curve normal to the
${\bf S}=$ constant surfaces; along any such trajectory 
we will introduce the
following matrices, $\theta$, $C(\beta)$, $D(\beta)$ and $G(\beta)$, 
each of which has its matrix element connecting two points 
$({\bf S}, \beta)$ and $({\overline{\bf S}}, \beta)$ along the same 
trajectory. The matrix element of $\theta$ is given by
\begin{eqnarray}\label{e4.28}
({\bf S}|\theta|{\bf {\overline S}})=
\theta({\bf S}-{\bf {\overline S}})=
\left\{\begin{array}{cc}
1 & ~~~~~~~~{\sf if}\hspace{4mm} 0 \leq {\bf {\overline S}} < {\bf S} \\
0 & ~~~~~~~~{\sf if} \hspace{4mm} 0 \leq {\bf S} < {\bf {\overline S}},  
\end{array}
\right. 
\end{eqnarray}
We now define $C$, $D$ and $G$ in matrix notations 
by (cf. (\ref{e3.13})-(\ref{e3.17}))
\begin{eqnarray}\label{e4.30}
C &\equiv& g^{-1} \theta h_{\bf S}^2,\\
D &\equiv& -2e^{-g {\bf S}}\theta e^{2 g{\bf S}}
\frac{h_{\bf S}}{h_{\beta}} \theta
e^{-g{\bf S}}
h_{\bf S}h_{\beta}
\end{eqnarray}
and
\begin{eqnarray}\label{e4.32}
G \equiv D(1+ T_{\beta} D)^{-1} = (1+ D T_{\beta} )^{-1}D,
\end{eqnarray}
in which the $\beta$-dependence of the matrices is not exhibited explicitly 
and $e^{\pm g{\bf S}}$, $h_{\bf S}$, $h_{\beta}$ are all 
diagonal matrices. (The conditions for the existence of inverse matrices,
like $(1+T_{\beta}D)^{-1}$ will be given later, after (\ref{e4.62}).)
For example, the matrix element of $C$ is
\begin{eqnarray}\label{e4.29}
({\bf S}|C|{\bf {\overline S}}) =
g^{-1}\theta({\bf S}-{\bf {\overline S}})
/({\bf \nabla}{\bf {\overline S}})^2.
\end{eqnarray}
Similar to (\ref{e3.13}) and (\ref{e3.15})-(\ref{e3.16}),
the product of any two matrices, say $A$ and $B$,
defined along the same trajectory (i.e., sharing the same $\beta$)
has its matrix element given by the usual rule
\begin{eqnarray}\label{e4.33}
({\bf S}|AB|{\bf {\overline S}}) =
\int_0^{\infty}d{\overline {\overline {\bf S}}}
({\bf S}|A|{\overline {\overline {\bf S}}})
( {\overline {\overline {\bf S}}}|B| {\overline {\bf S}}).
\end{eqnarray}.
The matrix element of its derivative 
$\frac{\partial}{\partial {\bf S}} AB$ is given by
\begin{eqnarray}\label{e4.34}
({\bf S}|\frac{\partial}{\partial {\bf S}}AB|{\bf {\overline S}})=
\int_0^{\infty}d{\overline {\overline {\bf S}}}
\frac{\partial}{\partial {\bf S}}
({\bf S}|A|{\bf {\overline {\overline S}}})
( {\overline {\overline {\bf S}}}|B|{\bf {\overline S}}),
\end{eqnarray}
with 
$\frac{\partial}{\partial {\bf S}}$ operating only on the left index ${\bf S}$,
whereas that of $\frac{\partial}{\partial \beta_j}AB$ is
\begin{eqnarray}\label{e4.35}
({\bf S}|\frac{\partial}{\partial \beta_j}AB|{\overline {\bf S}})=
({\bf S}|(\frac{\partial A}{\partial \beta_j})B + 
A(\frac{\partial B}{\partial \beta_j})|{\overline {\bf S}}).
\end{eqnarray}

\noindent
\underline{Theorem 2.}

\begin{eqnarray}\label{e4.36}
&1.~~g{\bf \nabla}{\bf S} \cdot {\bf \nabla} C = {\sf I},\\
&2.~~D~~{\sf and}~~G~~{\sf satisfy}\nonumber\\
&~~~~D = e^{-g{\bf S}} (1+CT_{\bf S})^{-1}C e^{g{\bf S}} =
e^{-g{\bf S}}C (1+T_{\bf S}C)^{-1}e^{g{\bf S}}\\
{\sf and}&\nonumber\\
&~~~~G = e^{-g{\bf S}} (1+CT)^{-1}C e^{g{\bf S}} =
e^{-g{\bf S}}C (1+TC)^{-1}e^{g{\bf S}}.
\end{eqnarray}
\hspace*{1cm}3. Furthermore, $D$ and $G$ are the Green's functions 
of $T_{\bf S}+V-E$ and $T+V-E$:
\begin{eqnarray}\label{e4.39}
(T_{\bf S} + V - E) D &=& {\sf I},\\
(H-E)G &=& (T+V-E)G = {\sf I},
\end{eqnarray}
where I denotes the unit matrix whose matrix element is
$\delta({\bf S}-{\overline {\bf S}})$.

\noindent
\underline{Proof} \\
1. For ${\bf S}>0$,
\begin{eqnarray}\label{e4.41}
\frac{\partial}{\partial {\bf S}} 
\theta ({\bf S}-\overline{\bf S}) = 
\delta ({\bf S}-\overline{\bf S}).
\end{eqnarray}
On account of the orthogonality 
${\bf \nabla}{\bf S}\cdot {\bf \nabla}\beta_j = 0$, we have 
\begin{eqnarray}\label{e4.42}
{\bf \nabla}{\bf S}\cdot {\bf \nabla} =
({\bf \nabla}{\bf S})^2 \frac{\partial}{\partial {\bf S}}.
\end{eqnarray}
Eq.(\ref{e4.36}) follows.\\ 
2. Define
\begin{eqnarray}\label{e4.43}
\overline{D} \equiv e^{g {\bf S}} D e^{-g{\bf S}}~~{\sf and}~~
\overline{G} \equiv e^{g {\bf S}} G e^{-g{\bf S}}.
\end{eqnarray}
Then, from (4.31) we see that
\begin{eqnarray}\label{e4.44}
{\overline D} = 
-2\theta e^{2g{\bf S}}\frac{h_{\bf S}}{h_{\beta}} \theta e^{-2g{\bf S}}
h_{\bf S} h_{\beta}, \\
\frac{\partial \overline{D}}{\partial {\bf S}} = 
-2 e^{2g {\bf S}} \frac{h_{\bf S}}{h_{\beta}} 
\theta e^{-2g{\bf S}} h_{\bf S} h_{\beta}, \\
\frac{\partial}{\partial {\bf S}} [\frac{h_{\beta}} {h_{\bf S}}
\frac{\partial \overline{D}}{\partial {\bf S}}] = 
-4 g e^{2g {\bf S}} \theta e^{-2g{\bf S}} 
h_{\bf S} h_{\beta} - 2 h_{\bf S} h_{\beta}.  
\end{eqnarray}
and, because of (4.22),
\begin{eqnarray}\label{e4.47}
T_{\bf S} \overline{D} = \frac{2 g}  {h_{\bf S} h_{\beta}} 
e^{2g {\bf S}} \theta e^{-2g{\bf S}} h_{\bf S} h_{\beta} + 1.
\end{eqnarray}
Multiplying by $C$ on the left, we find
\begin{eqnarray}
C T_{\bf S} \overline{D} = 2 \theta \frac{h_{\bf S}}{h_{\beta}} 
e^{2g {\bf S}} \theta e^{-2g{\bf S}} h_{\bf S} h_{\beta} + C =
- \overline{D} + C,\nonumber
\end{eqnarray}
and therefore
\begin{eqnarray}\label{e4.48}
(1+CT_{\bf S}) \overline{D} = C,
\end{eqnarray}
or
\begin{eqnarray}\label{e4.49}
\overline{D} = (1+CT_{\bf S})^{-1} C =
C(1+T_{\bf S}C)^{-1} .
\end{eqnarray}
Since according to (4.43)
\begin{eqnarray}\label{e4.50}
D = e^{-g {\bf S}} \overline{D} e^{g{\bf S}},
\end{eqnarray}
(4.37) is proved. 

By using (4.31), we derive
\begin{eqnarray}\label{e4.51}
\frac{\partial D}{\partial {\bf S}} = -gD
-2 e^{g {\bf S}} 
\frac{h_{\bf S}}{h_{\beta}} 
\theta e^{-g{\bf S}} h_{\bf S} h_{\beta}, 
\end{eqnarray}
\begin{eqnarray}\label{e4.52}
\frac{h_{\beta}} {h_{\bf S}}
\frac{\partial D}{\partial {\bf S}} = -g \frac{h_{\beta}}{h_{\bf S}}D 
-2 e^{g {\bf S}} \theta e^{-g{\bf S}} 
h_{\bf S} h_{\beta},  
\end{eqnarray}
and
\begin{eqnarray}\label{e4.53}
\frac{\partial}{\partial {\bf S}} (\frac{h_{\beta}} {h_{\bf S}}
\frac{\partial D}{\partial {\bf S}}) = 
- g 
\{[\frac{\partial}{\partial {\bf S}} (\frac{h_{\beta}} {h_{\bf S}})]
- g
\frac{h_{\beta}}{h_{\bf S}}\}D - 2 h_{\bf S} h_{\beta}.
\end{eqnarray}
On the other hand, because $\Phi = e^{-g{\bf S}}$ satisfies
$(T_{\bf S}+V-E)\Phi = 0$ and since
\begin{eqnarray}\label{e4.54}
T_{\bf S} \Phi =
- \frac {1}{2 h_{\bf S} h_{\beta}}
\frac{\partial}{\partial {\bf S}}
(\frac{h_{\beta}}{h_{\bf S}}\frac{\partial }{\partial {\bf S}}) 
e^{-g{\bf S}} = 
- \frac {1}{2 h_{\bf S} h_{\beta}}
\{-g [\frac{\partial}{\partial {\bf S}}
(\frac{h_{\beta}} {h_{\bf S}})] + g^2 
\frac{h_{\beta}} {h_{\bf S}}\} e^{-g{\bf S}},
\end{eqnarray}
we have
\begin{eqnarray}\label{e4.55}
\frac {g}{2 h_{\bf S} h_{\beta}}
\{[\frac{\partial}{\partial {\bf S}}
(\frac{h_{\beta}} {h_{\bf S}})] - g 
\frac{h_{\beta}} {h_{\bf S}}\} = -V+E.
\end{eqnarray}
Therefore (\ref{e4.53}) leads to
\begin{eqnarray}\label{e4.56}
-\frac {1}{2 h_{\bf S} h_{\beta}}
\frac{\partial}{\partial {\bf S}}
(\frac{h_{\beta}} {h_{\bf S}}
\frac{\partial D}{\partial {\bf S}}) +(V-E)D = 1;
\end{eqnarray}
i.e., $D$ satisfies (\ref{e4.39}) and is the Green's function of
$T_{\bf S}+V-E$.

From (\ref{e4.32}) and the identity
\begin{eqnarray}\label{e4.57}
1 = [1 - T_{\beta} D (1+ T_{\beta} D)^{-1}] (1+ T_{\beta} D)
\end{eqnarray}
we have 
\begin{eqnarray}\label{e4.58}
(T_{\bf S}+V-E)D = [1 - T_{\beta} D (1+ T_{\beta} D)^{-1}]
(1+ T_{\beta} D),
\end{eqnarray}
i.e.,
\begin{eqnarray}\label{e4.59}
(T_{\bf S}+V-E)D (1+ T_{\beta} D)^{-1} =
1 - T_{\beta} D (1+ T_{\beta} D)^{-1}.
\end{eqnarray}
From (\ref{e4.32}), $D (1+ T_{\beta} D)^{-1}$ is $G$; we derive
\begin{eqnarray}\label{e4.60}
(T+V-E)G = 1,
\end{eqnarray}
which establishes (4.40).

From $\overline{G} = e^{g {\bf S}} G e^{-g{\bf S}}$ and since $e^{g{\bf S}}$
commutes with $T_{\beta}$, it follows that
\begin{eqnarray}\label{e4.61}
\overline{G} = \overline{D} (1+T_{\beta}\overline{D})^{-1}.
\end{eqnarray}
By using $\overline{D} = C (1+T_{\bf S}C)^{-1}$, we have
\begin{eqnarray}
1+T_{\beta}\overline{D} &=& 1+T_{\beta}C(1+T_{\bf S}C)^{-1}\nonumber\\
                        &=& (1+T_{\bf S}C+T_{\beta}C)(1+T_{\bf S}C)^{-1}.
\nonumber
\end{eqnarray}
Its inverse is
\begin{eqnarray}
(1+T_{\beta}\overline{D})^{-1} = (1+T_{\bf S}C)(1+TC)^{-1}\nonumber
\end{eqnarray}
which leads to
\begin{eqnarray}\label{e4.62}
\overline{G} = \overline{D}(1+T_{\bf S}C)(1+TC)^{-1}=C(1+TC)^{-1},
\end{eqnarray}
and that gives (4.38). From the above expression, we see that 
\begin{eqnarray}
\overline{G}^{-1} = C^{-1} + T.~~~~~~~~~~~~~~~~~~~~~~~~~~~~~(4.38)'
\nonumber
\end{eqnarray}
Likewise, from (4.49) and (4.61), we can write
\begin{eqnarray}
\overline{D}^{-1} = C^{-1} + T_{\bf S}~~~~~~~~~~~~~~~~~~~~~~~~~~~
(4.37)' \nonumber
\end{eqnarray}
and
\begin{eqnarray}
\overline{G}^{-1} = \overline{D}^{-1} + T_{\beta}.~~~~~~~~~~~~~~~~~~~~~~~~~~~
(4.31)'\nonumber
\end{eqnarray}
These relations make transparent the connections between the original 
equations (4.38), (4.37) and (4.31).

To complete the proof, we have to examine the conditions, under which
$1+CT$, $1+CT_{\bf S}$ and  $1+DT_{\beta}$ have inverses.

As in (\ref{e3.20}), we assume $f_1 = f_1({\bf S}, \beta)$ to satisfy
\begin{eqnarray}\label{e4.63}
(1+CT)f_1 = 0.
\end{eqnarray}
Operating on its left by $-2g{\bf \nabla}{\bf S}\cdot {\bf \nabla}$, 
we derive
\begin{eqnarray}
-2g {\bf \nabla} {\bf S}\cdot {\bf \nabla}f_1 +
{\bf \nabla}^2 f_1 = 0.\nonumber
\end{eqnarray}
Hence
\begin{eqnarray}\label{e4.64}
{\bf \nabla} \cdot (e^{-2g{\bf S}} {\bf \nabla}f_1) = 0.
\end{eqnarray}
Multiplying by $f_1 d^N{\bf q} $ and integrating over all space, we have
\begin{eqnarray}\label{e4.65}
\int e^{-2g{\bf S}} ({\bf \nabla}f_1)^2 d^N{\bf q} = 0,
\end{eqnarray}
provided
\begin{eqnarray}\label{e4.66}
e^{-2g{\bf S}}f_1 {\bf \nabla}f_1 = 0~~~~~~~~{\sf at}~~~~~~\infty.
\end{eqnarray}
From (\ref{e4.65}), we see that the only solution is
${\bf \nabla}f_1 = 0$; i.e., $f_1 =$ constant. However, from the original
equation (\ref{e4.63}), it follows then $f_1 = 0$.
Consequently, $(1+CT)^{-1}$ exists provided (\ref{e4.66}) holds.

Next, we examine the equation
\begin{eqnarray}\label{e4.67}
(1+CT_{\bf S})f_2 = 0.
\end{eqnarray}
Operating $g {\bf \nabla}{\bf S}\cdot {\bf \nabla}$ on its left, 
we find
\begin{eqnarray}\label{e4.68}
g {\bf \nabla}{\bf S}\cdot {\bf \nabla}f_2 + T_{\bf S}f_2 = 0.
\end{eqnarray}
Multiplying (\ref{e4.68}) by $f_2h_{\bf S}h_{\beta}d{\bf S}$ and
integrating from ${\bf S} = 0$ to $\infty$, we have
\begin{eqnarray}\label{e4.69}
\int_0^{\infty} f_2\frac{\partial}{\partial {\bf S}}
(\frac{h_{\beta}} {h_{\bf S}} e^{-2g{\bf S}}
\frac{\partial f_2}{\partial {\bf S}})d{\bf S} = 0.
\end{eqnarray}
Assuming
\begin{eqnarray}\label{e4.70}
(f_2\frac{1}{h_{\bf S}}
\frac{\partial f_2}{\partial {\bf S}}) e^{-2g{\bf S}} h_{\beta} = 0
~~~~{\sf at}~~~~{\bf S}=0~~~~{\sf and}~~~~{\bf S} = \infty,
\end{eqnarray}
(which is the ${\bf S}$-component equivalence of (\ref{e4.66})), we derive
\begin{eqnarray}\label{e4.71}
\int_0^{\infty} e^{-2g{\bf S}}(\frac{1} {h_{\bf S}} 
\frac{\partial f_2}{\partial {\bf S}})^2 h_{\beta}h_{\bf S} d{\bf S}= 0.
\end{eqnarray}
Recognizing $d^N{\bf q} = h_{\bf S}h_{\beta}d{\bf S}d\beta$ is positive,
it follows then $\partial f_2/\partial {\bf S} = 0$. Substituting this
back to (\ref{e4.67}), we find
\begin{eqnarray}\label{e4.72}
f_2 = 0.
\end{eqnarray}
Hence, $(1+CT_{\bf S})^{-1}$ exists, provided (\ref{e4.70}) holds.

Lastly, we consider
\begin{eqnarray}\label{e4.73}
(1+DT_{\beta})f_3 = 0.
\end{eqnarray}
Applying $T_{\bf S}+V-E$ from the left and using (\ref{e4.39}), we obtain
\begin{eqnarray}\label{e4.74}
(T_{\bf S}+T_{\beta} +V-E)f_3 = 0.
\end{eqnarray}
Thus, the inverse of $(1+DT_{\beta})$ exists, provided we restrict
ourselves to a Hilbert space which excludes the ground-state of $H$.
Theorem~2 is then proved.

It is important to note that the existence of $(1+CT)^{-1}$, and
therefore $G$, as defined by (4.38), is free from the last
restriction.

\newpage

We return now to (\ref{e4.3})-(\ref{e4.6}). The following theorem will
enable us to derive the ground-state wave function 
$\Psi = e^{-g{\bf S}-\tau}$ of $H+\epsilon U$ in terms of
$e^{-g{\bf S}}$, the ground-state wave function of $H$:

\noindent
\underline{Theorem 3}
\begin{eqnarray}\label{e4.75}
e^{-\tau} = 1+(1+CT)^{-1} C \epsilon (-U+\Delta)e^{-\tau},
\end{eqnarray}
or equivalently,
\begin{eqnarray}\label{e4.76}
\Psi &=& e^{-g{\bf S}} + (1+DT_{\beta})^{-1} D \epsilon (-U+\Delta)\Psi
\nonumber\\
     &=& e^{-g{\bf S}} +G \epsilon (-U+\Delta)\Psi.
\end{eqnarray}

\noindent
\underline{Proof}. ~~Eq.(\ref{e4.11}) can also be written as
\begin{eqnarray}
(g {\bf \nabla}{\bf S}\cdot {\bf \nabla}+T)e^{-\tau} =
\epsilon (-U+\Delta)e^{-\tau},\nonumber
\end{eqnarray}
where $T = -\frac{1}{2} {\bf \nabla}^2$, as before. Since from
(\ref{e4.36}) $g{\bf \nabla}{\bf S}\cdot{\bf \nabla}C = 1$, and since
${\bf \nabla}(e^{-\tau}-1) = {\bf \nabla}e^{-\tau}$, the above expression
gives
\begin{eqnarray}\label{e4.77}
g {\bf \nabla}{\bf S}\cdot {\bf \nabla}(1+CT)(e^{-\tau}-1) =
\epsilon (-U+\Delta)e^{-\tau}.
\end{eqnarray}
Hence,
\begin{eqnarray}\label{e4.78}
e^{-\tau}-1 = (1+CT)^{-1}C \epsilon (-U+\Delta)e^{-\tau},
\end{eqnarray}
as can be derived by substituting (\ref{e4.78}) into (\ref{e4.77}). Thus
(\ref{e4.75}) follows. Multiplying (\ref{e4.78}) by $e^{-g{\bf S}}$ on
the left gives
\begin{eqnarray}\label{e4.79}
\Psi = e^{-g{\bf S}} + e^{-g{\bf S}}(1+CT)^{-1}C
\epsilon (-U+\Delta)e^{-\tau}
\end{eqnarray}
which leads to (\ref{e4.76}), on account of (\ref{e4.32}) and (4.38).
Theorem~3 is thereby established.

From (\ref{e4.75}), we may expand $e^{-\tau}$ as a power series of
$\epsilon$:
\begin{eqnarray}\label{e4.80}
e^{-\tau} = 1+ \epsilon(1+CT)^{-1}C (-U+\Delta) +
\epsilon^2 [(1+CT)^{-1}C (-U+\Delta)]^2 \nonumber\\
+ \cdots + \epsilon^n [(1+CT)^{-1}C (-U+\Delta)]^n + \cdots.
\end{eqnarray}
Likewise, from (\ref{e4.76}), we derive the perturbation series
\begin{eqnarray}\label{e4.81}
\Psi = \Psi_0 + \epsilon \Psi_1 + \epsilon^2 \Psi_2 + \cdots +
\epsilon^n \Psi_n + \cdots,
\end{eqnarray}
where
\begin{eqnarray}\label{e4.82}
\Psi_0 = e^{-g{\bf S}},
\end{eqnarray}
and for $n\geq 1$,
\begin{eqnarray}\label{e4.83}
\Psi_n = [G(-U+\Delta)]^n e^{-g{\bf S}},
\end{eqnarray}
with $G = (1+DT_{\beta})^{-1}D$, or
$e^{-g{\bf S}}(1+CT)^{-1}Ce^{g{\bf S}}$.

From (\ref{e4.83}), we have
\begin{eqnarray}\label{e4.84}
\Psi_n = G(-U+\Delta) \Psi_{n-1}.
\end{eqnarray}
Eqs. (\ref{e4.76}) and (\ref{e4.83}) can be directly established by using
$(H-E)G = 1$; i.e., $G$ being the Green's function of $(H-E)$. Likewise,
since
\begin{eqnarray}
(T_{\bf S} +V-E) e^{-g{\bf S}} = 0\nonumber
\end{eqnarray}
and
\begin{eqnarray}
(T_{\bf S} +V-E) \Psi = [-T_{\beta} + \epsilon (-U+\Delta)]\Psi,\nonumber
\end{eqnarray}
$\Psi$ also satisfies
\begin{eqnarray}\label{e4.85}
\Psi = e^{-g{\bf S}} + D(-T_{\beta} + \epsilon (-U+\Delta)] \Psi,
\end{eqnarray}
on account of (\ref{e4.39}), $D$ being the Green's function of
$(T_{\bf S} + V -E)$.

The new perturbation series (\ref{e4.81})-(\ref{e4.83}) is most effective, 
when it is complemented by an additional expansion in powers of $g^{-1}$,
the anharmonicity parameter. From
$C = g^{-1} \theta h_{\bf S}^2$
and (4.37)-(4.38), we see that the three Green's functions $C$, $D$ and 
$G$ are all $O(g^{-1})$. The n$^{th}$ order perturbative wave function 
$\epsilon^n \Psi_n$ is given by (\ref{e4.83}). By using (4.32) for $G$, 
$\Psi_n$ can be written as a power series in $D$, and therefore also in
$g^{-1}$:
\begin{eqnarray}\label{e4.86}
\Psi_n = \{[D-DT_\beta D + (DT_\beta)^2 D - \cdots + (-)^m (DT_\beta)^m D +
\cdots](-U+\Delta)\}^n e^{-g{\bf S}},
\end{eqnarray}
in which each $D$ assumes  the form  given by (4.31);  i.e.,
$$D = -2e^{-g {\bf S}}\theta e^{2 g{\bf S}}
\frac{h_{\bf S}}{h_{\beta}} \theta
e^{-g{\bf S}} h_{\bf S}h_{\beta},$$
a double definite integral along the trajectory. Through (4.38),
$\Psi_n$ can also be expressed as a different power series in $g^{-1}$ in 
terms of $C$:
\begin{eqnarray}\label{e4.87}
\Psi_n = e^{-g{\bf S}} \{[C-CTC + \cdots + (-)^m (CT)^mC +
\cdots](-U+\Delta)\}^n,
\end{eqnarray}
in which each $C$ is a single definite integral 
along the same trajectory. The equivalence between these two 
expressions rests on (4.37), which gives 
$$D = e^{-g{\bf S}} [C - CT_{\bf S}C + \cdots + (-)^m (CT_{\bf S})^mC +
\cdots ] e^{g{\bf S}}.$$

The identity between this power series expression  of  $D$ and  its 
double  integral form given  above has its origin in 
the Sturm-Liouville construction of the Green's function for a 
one-dimensional Schroedinger equation. Recalling that  $e^{-g{\bf S}}$ 
is a solution of the second order ordinary differential 
equation in ${\bf S}$,
\begin{eqnarray}\label{e4.88}
(T_{\bf S}+V-E)e^{-g{\bf S}} = 0,
\end{eqnarray}
we can form an irregular solution $F$ which satisfies the same equation
\begin{eqnarray}\label{e4.89}
(T_{\bf S}+V-E)F = 0
\end{eqnarray}
with 
\begin{eqnarray}\label{4.90}
F({\bf S}) =  e^{-g {\bf S}}\int\limits_0^{\bf S} e^{2g {\overline{\bf S}}}
(\frac{h_{\bf S}}{h_{\beta}})_{{\sf at}~{\overline{\bf S}}}  
d{\overline{\bf S}}, 
\end{eqnarray}
in which  the angular variable $\beta$ is kept fixed and  not  exhibited
explicitly in the argument.
The Green's function $D$ is related to the Wronskian-type expression by
\begin{eqnarray}\label{e4.91}
({\bf S}|D|\overline{\bf S}) = 
2 [e^{-g{\bf S}} F(\overline{\bf S}) - F({\bf S})
e^{-g{\overline {\bf S}}}]
(h_{\bf S} h_{\beta})_{{\overline {\bf S}},\beta}
\theta({\bf S} -\overline {\bf S}).
\end{eqnarray}
In ref.\cite{FLZ}, we used a one-dimensional example to demonstrate how this
form of the Green's function can enable us to evaluate the multi-instanton
expansion.
Here we have extended the application of a
one-dimensional Green's function to an $N$-dimensional problem.

\noindent
\underline{Theorem 4.} The energy shift $\epsilon\Delta$ is given by
\begin{eqnarray}\label{e4.92}
\epsilon\Delta = 
\frac{\int e^{-g{\bf S}}\epsilon U \Psi d^N {\bf q}}
{\int e^{-g{\bf S}} \Psi d^N {\bf q}}
\end{eqnarray}
where $d^N{\bf q} = h_{\bf S}h_{\beta}d{\bf S} d\beta$, as in (\ref{e4.26}) 
and the integration is extended over all space.

\noindent
\underline{Proof}~~~ From (4.31) or (4.91), the matrix element of $D(\beta)$
is
\begin{eqnarray}\label{e4.93}
({\bf S}|D(\beta)|\overline{\bf S}) = 
-2 e^{-g{\bf S}} \int_0^{\bf S}d
{\overline {\overline {\bf S}}}
(\frac{h_{\bf S}}{h_{\beta}})
_{{\overline {\overline {\bf S}}},\beta} \theta(
{\overline {\overline {\bf S}}}-{\overline {\bf S}})\cdot
e^{2g{\overline {\overline {\bf S}}}}
e^{-g{\overline {\bf S}}}
(h_{\bf S} h_{\beta})_{{\overline {\bf S}},\beta},
\end{eqnarray}
where the subscripts 
$\overline {\overline {\bf S}},\beta$ and
$\overline {\bf S},\beta$ indicate the arguments of the functions inside 
the parenthesis.

Let $\chi$ denotes the column matrix
\begin{eqnarray}
\chi ({\bf S},\beta) \equiv [-T_{\beta} + \epsilon (-U+\Delta)]
\Psi ({\bf S},\beta).\nonumber
\end{eqnarray}
Eq.(\ref{e4.85}) becomes
\begin{eqnarray}
\Psi = e^{-g{\bf S}} + D \chi.\nonumber
\end{eqnarray}
As $ {\bf S} \rightarrow \infty$, the left side should be zero, 
but the right side
is, on account of (\ref{e4.93}), controlled  by the
$\overline {\overline {\bf S}}$-integration in the upper region when 
$\overline {\overline {\bf S}} $ is near $ {\bf S} $; i.e., writing $D\chi$ 
as the product of $-2e^{-g{\bf S}}$ times
\begin{eqnarray}
\int_0^{\bf S}d{\overline {\overline {\bf S}}}
e^{2g{\overline {\overline {\bf S}}}}
(\frac{h_{\bf S}}{h_{\beta}})
_{{\overline {\overline {\bf S}}},\beta}
\int_0^{\overline {\overline {\bf S}}}d{\overline {\bf S}}
e^{-g{\overline {\bf S}}}
(h_{\bf S} h_{\beta})_{{\overline {\bf S}},\beta}
\chi(\overline {\bf S},\beta),\nonumber
\end{eqnarray}
we see that the factor to the right of (and including) the 
$\overline{\bf S}-$integration sign approaches $ e^{2g {\bf S}}$ times
\begin{eqnarray}\label{e4.94}
\int_0^{\infty}d{\overline {\bf S}}
e^{-g{\overline {\bf S}}}
(h_{\bf S} h_{\beta})_{{\overline {\bf S}},\beta}
\chi(\overline {\bf S},\beta),
\end{eqnarray}
which would make $D\chi \rightarrow \infty$
unless (\ref{e4.94}) is zero. Thus, the convergence of $\Psi$ at $\infty$
requires
\begin{eqnarray}\label{e4.95}
\int_0^{\infty}d{\bf S} e^{-g{\bf S}}
h_{\bf S} h_{\beta}
[-T_{\beta} +\epsilon (-U+\Delta)]\Psi = 0
\end{eqnarray}
at all $\beta$. Integrating over the angular variables
$d\beta = \prod_{j=1}^{N-1} d\beta_j,$ 
and because of
\begin{eqnarray}\label{e4.96}
-h_{\bf S} h_{\beta} T_{\beta} = \frac{1}{2}
\sum_{j=1}^{N-1}
\frac{\partial}{\partial \beta_j}
\frac{h_{\bf S} h_{\beta}}{h_{\beta_j}^2}
\frac{\partial}{\partial \beta_j}
\end{eqnarray}
and
$$\int d\beta h_{\bf S} h_{\beta} T_{\beta}\Psi = 0$$
we derive (\ref{e4.92}). This result is well known. (Since
$(H+\epsilon U)\Psi = (E+\epsilon \Delta)\Psi$, the multiplication of
$e^{-g{\bf S}}$ on the left and integrating over all space gives
immediately (\ref{e4.92}).) We have demonstrated that the same result can
also be derived, though somewhat awkwardly, by the Green's function
method developed in this paper.

The usual perturbation series requires to each order $n$, either a
product of infinite sums over all excited levels of the unperturbed
Hamiltonian $H$, or a sum over all possible paths through Feynman's
path integration method. The perturbation series formula derived
here is different; it depends only on quadratures along a single
trajectory, as will  be illustrated by examples in the next section. 

\newpage

\underline{Remarks.} Before leaving this section, we wish to address 
the connection between
the perturbation series developed in this section and the previous series
expansion given by (\ref{e3.28})-(\ref{e3.29}). In sections 1-3, we
discuss the relation between the Schroedinger wave function
$$e^{-g{\bf S}} = e^{-g{\bf S}_0 - \sigma}$$
and the classical Hamilton's action integral ${\bf S}_0 $, which satisfies
$$g^2( {\bf \nabla} {\bf S}_0)^2 = 2V = 2g^2v.$$
Thus formally we may regard $e^{-g{\bf S}_0}$ as the solution of a
Schroedinger-like wave equation whose eigenvalue happens to be $0$:
\begin{eqnarray}\label{e4.130}
(-\frac{1}{2}{\bf \nabla}^2 + V_c)e^{-g{\bf S}_0} = 0
\end{eqnarray}
where 
\begin{eqnarray}\label{e4.131}
V_c = g^2 v -\frac{1}{2}g {\bf \nabla}^2 {\bf S}_0.
\end{eqnarray}
Since $e^{-g{\bf S}}$ satisfies
\begin{eqnarray}\label{e4.132}
(-\frac{1}{2}{\bf \nabla}^2 + V_c +
\frac{1}{2} g{\bf \nabla}^2 {\bf S}_0)e^{-g{\bf S}} =
E e^{-g{\bf S}},
\end{eqnarray}
we may treat (\ref{e4.130}) as the unperturbed Schroedinger equation, and
(\ref{e4.132}) as the perturbed one, with
$\frac{1}{2} g{\bf \nabla}^2 {\bf S}_0$ as the perturbative Hamiltonian
and $E$ as the energy shift. The corresponding small $\epsilon$-parameter
that characterize the perturbation becomes $g^{-1}$, since
$\frac{1}{2} g{\bf \nabla}^2 {\bf S}_0$ is $O(g^{-1})$ times $V_c$. This
explains the similarity between (\ref{e3.28}) for $e^{-\sigma}$ and
(\ref{e4.75}) for $e^{-\tau}$; it also identifies the $g^{-1}$-expansion
with the perturbative $\epsilon$-expansion.

\newpage

\section*{\bf {5. Examples}}
\setcounter{section}{5}
\setcounter{equation}{0}

From (\ref{e4.32}) and (4.38), the Green's function $G$ can be either
expressed in terms of the single integral form $C$, or equivalently, the
double integral form $D$:
\begin{eqnarray}\label{e5.1}
G = e^{-g{\bf S}} (1+CT)^{-1}C e^{g{\bf S}}
\end{eqnarray}
or
\begin{eqnarray}\label{e5.2}
G = (1+ D T_{\beta} )^{-1}D.
\end{eqnarray}
These two complement each other in their applications. As in 
(\ref{e4.8}) - (\ref{e4.9}), let $e^{-g{\bf S}}$ be the ground-state wave
function of $H$. The corresponding ground-state wave function
\begin{eqnarray}\label{e5.3}
\Psi = e^{-g{\bf S}-\tau}
\end{eqnarray}
of $H+\epsilon U$ can also be expressed either in terms of $C$
(eq.(\ref{e4.75})):
\begin{eqnarray}\label{e5.4}
e^{-\tau} = 1+(1+CT)^{-1} C \epsilon (-U+\Delta)e^{-\tau},
\end{eqnarray}
or in terms of $D$ (eq.(\ref{e4.76})):
\begin{eqnarray}\label{e5.5}
\Psi = e^{-g{\bf S}} + (1+DT_{\beta})^{-1} D \epsilon (-U+\Delta)\Psi.
\end{eqnarray}
As shown in  (\ref{e4.80}) and (\ref{e4.86}), both can be used for the
perturbation series expansion.
In this section, we shall examine several examples which illustrate
the different merits of these two formulations.

For clarity, consider the one-dimensional case, in which
(\ref{e5.2}) and (\ref{e5.5}) reduce to $G=D$ and
\begin{eqnarray}\label{e5.6}
\Psi = e^{-g{\bf S}} + D \epsilon (-U+\Delta)\Psi.
\end{eqnarray}
As in (\ref{e4.43}), introduce
\begin{eqnarray}
\overline{D} \equiv e^{g {\bf S}} D e^{-g{\bf S}},\nonumber
\end{eqnarray}
which, in accordance with (\ref{e4.48}) and since $T_{\bf S}=T$ in the
one-dimensional case, is related to $C$ by
\begin{eqnarray}\label{e5.7}
(1+CT) \overline{D} = C.
\end{eqnarray}
In terms of $\overline{D}$, (\ref{e5.6}) becomes
\begin{eqnarray}\label{e5.8}
e^{-\tau} = 1+ \overline{D} \epsilon (-U+\Delta)e^{-\tau}.
\end{eqnarray}

In order that the above expression is equivalent to the alternative form
(\ref{e5.4}) in terms of $C$, we need to convert (\ref{e5.7}) into
\begin{eqnarray}\label{e5.9}
\overline{D} = (1+CT)^{-1} C.
\end{eqnarray}
According to (\ref{e4.66}), this requires
\begin{eqnarray}\label{e5.10}
e^{-2g{\bf S}}f_1 {\bf \nabla}f_1 = 0
\end{eqnarray}
at the boundary, where, for (\ref{e5.8}) , the function $f_1$ is
\begin{eqnarray}\label{e5.11}
f_1 = \overline{D} (-U + \Delta) e^{-\tau},
\end{eqnarray}
and in one-dimension, the boundary consists of
\begin{eqnarray}\label{e5.12}
x = \pm \infty.
\end{eqnarray}
We recall further that $C$ satisfies (\ref{e4.36}) which is now an ordinary
first order differential equation
\begin{eqnarray}\label{e5.13}
g\frac{d{\bf S}}{dx}\frac{dC}{dx} = 1;
\end{eqnarray}
this determines $C$ up to an integration constant, which can be set by
requiring
\begin{eqnarray}\label{e5.14}
C = 0~~~~~~~~~{\sf at}~~~~~~{\bf S} = 0.
\end{eqnarray}
Now, $D$ is the solution of the second order differential equation
\begin{eqnarray}\label{e5.15}
(-\frac{1}{2}\frac{d^2}{dx^2} + V - E) D = 1,
\end{eqnarray}
and therefore contains two independent integration constants. In view of
(\ref{e5.9}) and (\ref{e5.14}), we require
\begin{eqnarray}\label{e5.16}
\overline{D} = 0~~~~~~~~{\sf at}~~~~~~~~~~{\bf S} = 0
\end{eqnarray}
and write for the right hand side of (\ref{e5.11})
\begin{eqnarray}\label{e5.17}
\overline{D} (-U + \Delta) e^{-\tau} = -2
\int\limits_0^x e^{2g{\bf S}(y)}dy
\int\limits_{-\infty}^y e^{-2g{\bf S}(z)}
(-U+\Delta)e^{-\tau(z)}dz
\end{eqnarray}
so that (\ref{e5.10}) holds at the boundary $x=-\infty$. (For convenience,
we choose ${\bf S} = 0$ at $x=0$.) In order to satisfy (\ref{e5.10}) at the
other boundary $x = \infty$, we set
\begin{eqnarray}\label{e5.18}
\int\limits_{-\infty}^{\infty} e^{-2g{\bf S}(x)-\tau(x)}
(-U+\Delta)dx = 0;
\end{eqnarray}
otherwise, (\ref{e5.17}) would $\sim e^{2g{\bf S}}$ at $x=\infty$, violating
(\ref{e5.10}) - (\ref{e5.11}). Thus, the energy shift $\Delta$ is also
determined, and (\ref{e5.18}) is a special case of (\ref{e4.92}).
On the other hand, if we use (\ref{e5.1}), expressing $G$ in terms of 
the single integral form $C$, then $\Delta$ would be determined by
differentiations at ${\bf S} = {\bf 0}$, similar to (\ref{e1.32}) and 
(1.34). In the following, we shall solve two examples by using $G$,
first in terms of $C$, then in terms of $D$.

\noindent
\underline{Example 1}.~~~~~Consider a one-dimensional harmonic oscillator 
with
\begin{eqnarray}\label{e4.97}
H = -\frac{1}{2} \frac{d^2}{dx^2} + \frac{g^2}{2} x^2
\end{eqnarray}
and a perturbative potential
\begin{eqnarray}\label{e4.98}
\epsilon U = \epsilon x^{2p},
\end{eqnarray}
where $p$ is a positive integer. The unperturbed ground-state wave function 
is
\begin{eqnarray}\label{e4.99}
\Phi = e^{-g S} = e^{- \frac{g}{2} x^2 }.
\end{eqnarray}
Since $ S =  \frac{1}{2} x^2$, we have
\begin{eqnarray}\label{e4.100}
h_S^2 = (dS/dx)^{-2} = x^{-2} = (2S)^{-1}.
\end{eqnarray}
For $n>0$ and in accordance with (\ref{e5.14}),
\begin{eqnarray}\label{e4.101}
Cx^{2n} = \int_0^S \frac{dS}{g2S}x^{2n} = \frac{1}{g(2n)} x^{2n}.
\end{eqnarray}

\noindent
\underline{Lemma 1} 
\begin{eqnarray}\label{e4.102}
(1+CT)^{-1}C[x^{2n} - \Gamma_{1~n}] = 
\sum_{m=1}^n \Gamma_{m~n}x^{2m}.
\end{eqnarray}
where 
\begin{eqnarray}\label{e4.103}
\Gamma_{n~n} &=& \frac{1}{g2n},~~~\Gamma_{n-1~n} = \frac{2n-1}{2g^2(2n-2)},
\nonumber\\
\Gamma_{m~n} &=& \frac{(2n-1)(2n-3)\cdots(2m+1)}{m(2g)^{n-m+1}}~~~~
{\sf for}~~~~1 \leq m \leq n-1 
\end{eqnarray}
and in particular
\begin{eqnarray}
\Gamma_{1~n} = \frac{(2n-1)!!}{(2g)^n}.\nonumber
\end{eqnarray}

\noindent
\underline{Proof}~~~~Since $T = -\frac{1}{2} \frac{d^2}{dx^2}$,
$$(1+CT)^{-1}C = C-CTC +(-CT)^2C + \cdots + (-CT)^mC + \cdots$$
and
$$-TCx^{2n} = \frac{1}{2g} (2n-1) x^{2n-2},$$
we find 
\begin{eqnarray}
-CTC x^{2n} = \frac{2n-1}{2g^2(2n-2)} x^{2n-2} = \Gamma_{n-1~n} x^{2n-2},
\nonumber\\
(-CT)^2C x^{2n} = \frac{(2n-1)(2n-3)}{2^2g^3(2n-4)} x^{2n-4} = 
\Gamma_{n-2~n} x^{2n-4},\nonumber
\end{eqnarray}
$\cdots$, until
\begin{eqnarray}
(-CT)^{n-1}C x^{2n} = \frac{(2n-1)!!}{(2g)^n} x^2 = 
\Gamma_{1~n} x^2.\nonumber
\end{eqnarray}
But, $(-CT)^nC x^{2n} = C\Gamma_{1~n} $, which by itself would be $\infty$.
On the other hand,
\begin{eqnarray}\label{e4.104}
(-CT)^nC x^{2n} -C\Gamma_{1~n} = 0;
\end{eqnarray}
Lemma 1 is then proved.

As in (\ref{e4.9}), the eigenstate of
\begin{eqnarray}
{\cal H} = -\frac{1}{2}\frac{d^2}{dx^2} + \frac{g^2}{2} x^2 + \epsilon x^{2p}
\nonumber
\end{eqnarray}
is $ e^{-\frac{1}{2}gx^2-\tau}$. Write
\begin{eqnarray}\label{e4.105}
e^{-\tau} = 1 + \sum_{l=1}^{\infty} a_l x^{2l}.
\end{eqnarray}
From (\ref{e4.75})
\begin{eqnarray}\label{e4.106}
\sum_{l=1}^{\infty} a_l x^{2l} = (1+CT)^{-1} C \epsilon 
(-x^{2p}+\Delta) (1 + \sum_{l=1}^{\infty} a_l x^{2l}).
\end{eqnarray}
By using (\ref{e4.102}) and (\ref{e4.105}), we derive
\begin{eqnarray}\label{e4.107}
\Delta = \Gamma_{1~p} + \sum_{l=1}^{\infty} a_l (\Gamma_{1~l+p} -
\Delta \Gamma_{1~l})
\end{eqnarray}
and for $n \geq 1$
\begin{eqnarray}\label{e4.108}
a_n = -\epsilon \Gamma_{n~p} -\epsilon \sum_{l=1}^{\infty} a_l 
(\Gamma_{n~l+p} - \Delta \Gamma_{n~l}).
\end{eqnarray}
where $\Gamma_{m~n} = 0$ for $m > n$; otherwise it is given by 
(\ref{e4.103}). From (\ref{e4.107}),we see that
\begin{eqnarray}\label{e4.109}
\epsilon \Delta = -a_1.
\end{eqnarray}
Expand 
\begin{eqnarray}
&\epsilon \Delta = \epsilon \Delta (1) + \epsilon^2 \Delta (2) +
\epsilon^3 \Delta (3) + \cdots \nonumber\\
{\sf and}~~~~~~~~~~~~~~& \nonumber\\
&a_n = \epsilon a_n(1) + \epsilon^2 a_n(2) + \epsilon^3 a_n(3)  + \cdots;
\nonumber
\end{eqnarray}
it follows then
\begin{eqnarray}\label{e4.110}
\Delta (1) = \Gamma_{1~p},~~~~~~~~~~~~a_n(1) = -\Gamma_{n~p},\nonumber \\
\Delta (2) = \sum_{m=1}^{p} \Gamma_{m~p} (-\Gamma_{1~m+p} +
\Gamma_{1~m} \Gamma_{1~p}) \\
a_n(2) =  \sum_{m=1}^{p} \Gamma_{m~p} (\Gamma_{n~m+p} -
\Gamma_{n~m} \Gamma_{1~p}), \nonumber
\end{eqnarray}
etc.. (Because of (\ref{e4.108}), $a_n(2) = 0$ for $n > 2p$.)

As a special case, for
\begin{eqnarray}\label{e4.111}
\epsilon U = \epsilon x^4
\end{eqnarray}
and therefore $p = 2$, we find
\begin{eqnarray}\label{e4.112}
\epsilon \Delta (1) = \frac{3 \epsilon}{4 g^2},~~~~~
\epsilon^2 \Delta (2) =  -\frac{21 \epsilon^2}{8 g^5},
\end{eqnarray}
the same as the usual perturbation results.

\noindent\underline{Example~2}~~~~Using the same one-dimensional 
harmonic oscillator Hamiltonian (\ref{e4.97}) as the unperturbed 
$H$, we consider now an odd power perturbation potential
\begin{eqnarray}\label{e4.113}
\epsilon U = \epsilon x^{2p+1}.
\end{eqnarray}
The analysis given below is parallel to that in the first example,
but with some changes, as will be indicated.

For $n \geq 0$,
\begin{eqnarray}\label{e4.114}
Cx^{2n+1} = \int_0^S \frac{dS}{g2S}x^{2n+1} = \frac{1}{g(2n+1)} x^{2n+1}.
\end{eqnarray}
It is straightforward to establish the following:

\noindent\underline{Lemma~2}
\begin{eqnarray}\label{e4.115}
(1+CT)^{-1}Cx^{2n+1} = \sum_{m=0}^n \gamma_{m~n}x^{2m+1}
\end{eqnarray}
where 
\begin{eqnarray}\label{e4.116}
\gamma_{n~n} = \frac{1}{g(2n+1)},~~~\gamma_{n-1~n} = \frac{n}{g^2(2n-1)},
\nonumber\\
\gamma_{m~n} = \frac{n(n-1)\cdots(m+1)}{g^{n-m+1}(2m+1)}~~~~
{\sf for}~~~~0 \leq m < n 
\end{eqnarray}
and in particular
$$\gamma_{0~n} = \frac{n!}{g^{n+1}}.$$
Instead of (\ref{e4.105}), we now set
\begin{eqnarray}\label{e4.117}
e^{-\tau} = 1 + \sum_{n=1}^{\infty} b_n x^{n}.
\end{eqnarray}
From (\ref{e4.75}),
\begin{eqnarray}\label{e4.118}
\sum_{n=1}^{\infty} b_n x^{n} = (1+CT)^{-1} C \epsilon 
(-x^{2p+1}+\Delta) (1 + \sum_{n=1}^{\infty} b_n x^{n}).
\end{eqnarray}

It is convenient to extend the definitions of $\gamma_{m~n}$
and $\Gamma_{m~n}$ (as in (\ref{e4.103})), by defining
\begin{eqnarray}\label{e4.119}
\gamma_{m~n}=\Gamma_{m~n}=0~~~~~~{\sf~for}~~~~m>n
\end{eqnarray}
and
\begin{eqnarray}
\Gamma_{0~n}=0.\nonumber
\end{eqnarray}
Using the two Lemmas, we derive
\begin{eqnarray}\label{e4.120}
\Delta = \sum_{l} (b_{2l+1}~\Gamma_{1~l+p+1} -
\Delta~b_{2l}~\Gamma_{1~l}),\nonumber\\
b_{2n} = -\epsilon \sum_{l} 
(b_{2l+1}~\Gamma_{n~l+p+1} - \Delta~b_{2l}~\Gamma_{n~l})
\end{eqnarray}
and
$$b_{2n+1} = -\epsilon \gamma_{n~p} -\epsilon \sum_{l} 
(b_{2l}~\gamma_{n~l+p} - \Delta~b_{2l+1}~\gamma_{n~l}).$$
Here and in the following, whenever the range of the summation index is
not exhibited, it reads
\begin{eqnarray}\label{e4.121}
l~({\sf~or}~m,~n)=0,~1,~2,\cdots,~\infty.
\end{eqnarray}
By comparing the first two equations in (\ref{e4.120}), we have 
\begin{eqnarray}\label{e4.122}
\epsilon \Delta = b_2.
\end{eqnarray}
It can be readily verified that the following power series expansions
hold:
\begin{eqnarray}\label{e4.123}
\epsilon \Delta = \epsilon^2 \Delta (2) + \epsilon^4 \Delta (4) +
\epsilon^6 \Delta (6) + \cdots \nonumber\\
b_{2n} = \epsilon^2 b_{2n}(2) + \epsilon^4 b_{2n}(4) + 
\epsilon^6 b_{2n}(6) + \cdots,\\
b_{2n+1} = \epsilon b_{2n+1}(1) + \epsilon^3 b_{2n+1}(3) + 
\epsilon^5 b_{2n+1}(5) + \cdots,\nonumber 
\end{eqnarray}
in which

\begin{eqnarray}\label{e4.124}
b_{2n+1}(1) = -\gamma_{n~p},~~~~~~~
\Delta (2) = -\sum_l \gamma_{l~p} \Gamma_{1~l+p+1},\nonumber\\ 
b_{2n}(2) =  \sum_l \gamma_{l~p} \Gamma_{n~l+p+1},\nonumber\\
b_{2n+1}(3) = -\sum_{l,m} \gamma_{l~p} \Gamma_{m~l+p+1}\gamma_{n~m+p} -
\Delta (2)\sum_m \gamma_{n~m}\gamma_{m~p},\\
\Delta (4) = -\Delta (2)\sum_{l,m} 
(\gamma_{l~p}\gamma_{m~p}\Gamma_{1~l+p+1} +
\Gamma_{l~m+p+1}\gamma_{m~p}\Gamma_{1~l}) \nonumber\\
- \sum_{l,m,n} \gamma_{l~p}\Gamma_{m~l+p+1}
\gamma_{n~m+p}\gamma_{1~n+p+1},\nonumber
\end{eqnarray}
etc.

In  the special case of $p=0$, we have 
\begin{eqnarray}\label{e4.125}
\epsilon U = \epsilon x,
\end{eqnarray}
the above expressions lead to
\begin{eqnarray}\label{e4.126}
\epsilon^2 \Delta(2) =-\frac{ \epsilon^2}{2g^2} 
\end{eqnarray}
and all other $\Delta(2n)=0$. This confirms the exact result, since
\begin{eqnarray}\label{e4.127}
V+\epsilon U = \frac{g^2}{2} x^2 + \epsilon x = 
\frac{g^2}{2}(x+\frac{\epsilon}{g^2})^2 -\frac{\epsilon^2}{2g^2}. 
\end{eqnarray}
The  lowest eigenvalue of $V$ is $\frac{g}{2}$, and that of $V+\epsilon U$
is $\frac{g}{2}-\frac{\epsilon^2}{2g^2}$. 
Likewise, (\ref{e4.123})-(\ref{e4.124}) yield
\begin{eqnarray}\label{e4.128}
b_1 = \epsilon b_1(1) = -\epsilon \gamma_{0~0} =-\frac{ \epsilon}{g}, 
\nonumber\\
b_2 = \epsilon^2 b_2(2) = \epsilon^2 \gamma_{0~0} \Gamma_{1~1} =
-\frac{ \epsilon^2}{2g^2}, \\
b_3 = \epsilon^3 b_3(3) = -\epsilon^3 \gamma_{0~0} \gamma_{1~1}\Gamma_{1~1} =
-\frac{ \epsilon^3}{6g^3}, \nonumber
\end{eqnarray}
etc., so that
\begin{eqnarray}
e^{-\tau} = 1 + \sum_{n=1}^{\infty} b_n x^{n} = e^{-\epsilon/g}
\nonumber
\end{eqnarray}
which leads to the expected expression 
\begin{eqnarray}\label{e4.129}
e^{-g{\bf S}-\tau} \propto
e^{-\frac{g}{2}(x+\frac{\epsilon}{g^2})^2}.
\end{eqnarray}

So far we have only used the Green's function $G$ in terms of the single 
integral form $C$. Next, we will repeat the same examples, but using the 
double integral form $D$.

\noindent
\underline{Example 1} (repeat) ~~~We return to the same simple harmonic oscillator
problem (\ref{e4.97}) - (\ref{e4.98}), with 
$$H = -\frac{1}{2}\frac{d^2}{dx^2} + \frac{g^2}{2} x^2$$
and the perturbation
\begin{eqnarray}\label{e5.19}
\epsilon U = \epsilon x^{2p}.\nonumber
\end{eqnarray}
In order to derive the perturbation series (\ref{e4.105}) - (\ref{e4.110})
for $e^{-\tau}$ and $\Delta$ using $D$, we have to deal with integrals of the type
\begin{eqnarray}\label{e5.20}
\overline{D} x^{2n} = -2 \int\limits_0^x e^{gy^2}dy
\int\limits_{-\infty}^y e^{-gz^2} z^{2n} dz.
\end{eqnarray}
Introduce
\begin{eqnarray}
\xi = \sqrt{g}x,~~~ \eta = \sqrt{g}y,~~~ \zeta = \sqrt{g}z.\nonumber
\end{eqnarray}
(\ref{e5.20}) becomes
\begin{eqnarray}\label{e5.21}
\overline{D} x^{2n} = -\frac{2}{g^{n+1}} \int\limits_0^{\xi} e^{\eta^2}
d\eta \int\limits_{-\infty}^{\eta} e^{-\zeta^2} \zeta^{2n} d\zeta.
\end{eqnarray}
Let $H_l(\zeta)$ be the usual Hermite polynomials, obtained by the
generating function
\begin{eqnarray}\label{e5.22}
e^{-t^2+2t\zeta} = \sum\limits_{l=0}^{\infty} \frac{t^l}{l!} H_l(\zeta).
\end{eqnarray}
The integrand in (\ref{e5.21}) can be expressed in terms of $H_l(\zeta)$
through
\begin{eqnarray}\label{e5.23}
\zeta^{2n} = \sum\limits_{m=0}^{n} \frac{(2n)!}{2^{2n}m!(2n-2m)!}
H_{2n-2m}(\zeta).
\end{eqnarray}
For $l \geq 1$,
\begin{eqnarray}
e^{-\xi^2} H_l(\xi) = -\frac{d}{d\xi} [e^{-\xi^2} H_{l-1}(\xi)]; \nonumber
\end{eqnarray}
therefore
\begin{eqnarray}\label{e5.24}
\int\limits_0^{\xi} e^{\eta^2} d\eta \int\limits_{-\infty}^{\eta}
e^{-\zeta^2} H_l(\zeta) d\zeta = -\int\limits_0^{\xi} H_{l-1}(\eta) d\eta
\nonumber \\
~~~~~~~~~~~~~~~~~~= -\frac{1}{2l} [H_l(\xi)-H_l(0)]
\end{eqnarray}
and
\begin{eqnarray}\label{e5.25}
\overline{D} H_l(\sqrt{g} x) = -\frac{2}{g}
\int\limits_0^{\xi} e^{\eta^2} d\eta \int\limits_{-\infty}^{\eta}
e^{-\zeta^2} H_l(\zeta) d\zeta
\nonumber \\
~~~~~~~~~~~~~~~~~~= \frac{1}{lg} [H_l(\xi)-H_l(0)].
\end{eqnarray}
Since $H_0(\zeta) = 1$, by separating out in (\ref{e5.23}) the term
$m=n$ in the sum, we have
\begin{eqnarray}\label{e5.26}
\zeta^{2n} = g^n \Gamma_{1~n} +
\sum\limits_{m=0}^{n-1} \frac{(2n)!}{2^{2n}m!(2n-2m)!}
H_{2n-2m}(\zeta),
\end{eqnarray}
where
\begin{eqnarray}
\Gamma_{1~n} = \frac{(2n-1)!!}{(2g)^n}, \nonumber
\end{eqnarray}
the same as (\ref{e4.103}). From (\ref{e5.25}) and (\ref{e5.26}), it follows
that
\begin{eqnarray}\label{e5.27}
\overline{D} (x^{2n} - \Gamma_{1~n}) =
\frac{(2n)!}{(4g)^n} \frac{1}{g}
\sum\limits_{l=1}^{n} \frac{1}{(n-l)!(2l)!2l}
[H_{2l}(\sqrt{g} x) - H_{2l}(0)].
\end{eqnarray}

The identity between the above expression and $(1+CT)^{-1}C(x^{2n}-
\Gamma_{1~n})$ given by (\ref{e4.102}) can be established by first
applying $T$ onto (\ref{e5.25}). Through the well known formula
\begin{eqnarray}\label{e5.28}
\frac{d^2}{d\xi^2} H_l(\xi) = 2 \xi \frac{d}{d\xi} H_l(\xi) - 2l H_l(\xi),
\end{eqnarray}
we find
\begin{eqnarray}\label{e5.29}
T \overline{D} H_l(\sqrt{g} x) = \frac{1}{l}
[-x\frac{d}{dx} H_l(\sqrt{g} x) + l H_l(\sqrt{g} x)].
\end{eqnarray}
Next, apply $C$ onto (\ref{e5.29}):
\begin{eqnarray}\label{e5.30}
C T \overline{D} H_l(\sqrt{g} x) &=& \frac{1}{g} \int\limits_0^x
\frac{dx}{x} T \overline{D} H_l(\sqrt{g} x)\nonumber\\
&=& -\frac{1}{lg}[H_l(\sqrt{g} x) - H_l(0)] + C H_l(\sqrt{g}x).
\end{eqnarray}
Combining (\ref{e5.25}) and (\ref{e5.30}), we obtain
\begin{eqnarray}\label{e5.31}
(1+CT) \overline{D} H_l(\sqrt{g} x) = C H_l(\sqrt{g} x).
\end{eqnarray}
Hence, the multiplication of $(1+CT)$ and (\ref{e5.27}) leads to
\begin{eqnarray}\label{e5.32}
(1+CT) \overline{D} (x^{2n} - \Gamma_{1~n}) = C (x^{2n} - \Gamma_{1~n}),
\end{eqnarray}
which gives the equality between (\ref{e5.27}) and (\ref{e4.102}).
Thus, both forms of $G$, (\ref{e5.1}) and (\ref{e5.2}), yield the same
perturbation series expansion  (\ref{e4.109}) - (\ref{e4.110}).

It is instructive to note that in (\ref{e4.102}),
$$(1+CT)^{-1}C[x^{2n} - \Gamma_{1~n}] = 
\sum_{m=1}^n \Gamma_{m~n}x^{2m},$$
each term, $(1+CT)^{-1}Cx^{2n}$ or $(1+CT)^{-1}C\Gamma_{1~n}$, is by itself 
$\infty$, which necessitates the combination that appears on the
left-hand side. Using the single-integral form $C$ for $G$, the energy shift
$\Delta$ is required to make the integral finite at $x=0$. On the other
hand, $\overline{D}x^{2n}$, or 
$$\overline{D}\Gamma_{1~n} = -2 \int\limits_0^x e^{gy^2} dy 
\int\limits_{-\infty}^y e^{-gz^2}\Gamma_{1~n} dz,$$
is well-defined at any finite $x$. As $x\rightarrow \infty$, either 
$\overline{D}x^{2n}$ or
\begin{eqnarray}\label{e5.65}
\overline{D}\Gamma_{1~n} \sim -2 [\Gamma_{1~n}
\int\limits_{-\infty}^{\infty} e^{-gz^2} dz ] x e^{gx^2}  
\end{eqnarray}
approaches $\infty$, which, when multiplied by $e^{-\frac{1}{2}gx^2}$, 
would lead to a perturbed wave function $e^{-g{\bf S}-\tau}$ divergent at
$x=\infty$. As shown in (\ref{e5.17}), this requires the energy shift 
$\Delta$ to be determined by (\ref{e5.18}); in this example, it is 
equivalent to have the subtraction $-\overline{D}\Gamma_{1~n}$ in
(\ref{e5.27}).

\underline{Example 2} (repeat). To complete the one-dimensional analysis, 
we continue with the same simple harmonic oscillator $H$ as the 
unperturbed Hamiltonian, but with the perturbation given by
\begin{eqnarray}\label{e5.66}
\epsilon U = \epsilon x^{2p+1}.
\end{eqnarray}
Since by repeated partial integrations,
\begin{eqnarray}
\int\limits_{-\infty}^y e^{-gz^2}z^{2n+1} dz &=& - \frac{1}{2g}
[y^{2n} + \frac{n}{g} y^{2(n-1)} + \nonumber\\
&&\frac{n(n-1)}{g^2}y^{2(n-2)} + \cdots + \frac{n!}{g^n}] e^{-gy^2},\nonumber
\end{eqnarray}
we find
\begin{eqnarray}
\overline{D} x^{2n+1} &=& -2 \int\limits_0^x e^{gy^2} dy
\int\limits_{-\infty}^y e^{-gz^2}z^{2n+1} dz \nonumber\\
&=& \sum\limits_{m=0}^{n} \gamma_{m~n} x^{2m+1},\nonumber
\end{eqnarray}
the same as (\ref{e4.115}) for $(1+CT)^{-1} C x^{2n+1}$, with $\gamma_{m~n}$
given by (\ref{e4.116}). It is straight forward to see that using either 
$\overline{D}$, or $(1+CT)^{-1} C$, we can arrive at the same perturbation 
series (\ref{e4.123}) and (\ref{e4.124}).

\newpage
 
\noindent
\section*{\bf 6. Excited States}
\setcounter{section}{6}
\setcounter{equation}{0}

In this section, we apply the new approach to excited states.

Let $ e^{-g {\bf S}} $ be the ground-state of (1.1) 
$$H e^{-g {\bf S}} = E e^{-g {\bf S}}$$
where
$$H = -\f{1}{2}\nabla^2+g^2v$$
and, as in (1.4)--(1.5),
\begin{eqnarray}
v ({\bf q})\geq 0.
\end{eqnarray}
In this section, we assume $ e^{-g {\bf S}} $ to be already known; our 
purpose is to derive the excited states $ \Psi_{ex} $, which satisfies
\begin{eqnarray}
H\Psi_{ex}=(E+{\cal E})\Psi_{ex}
\end{eqnarray}
Write
\begin{eqnarray}
\Psi_{ex}=\chi({\bf q})e^{-g {\bf S}}.
\end{eqnarray}
Since
$$
{\bf \nabla}^2\Psi_{ex}=({\bf \nabla}^2 \chi - 
2g{\bf \nabla} \chi \cdot {\bf \nabla} {\bf S})e^{-g {\bf S}}
+\chi ({\bf \nabla}^2 e^{-g {\bf S}}),
$$
(6.1)-(6.3) lead to
\begin{eqnarray}
g{\bf \nabla} {\bf S} \cdot {\bf \nabla} \chi - 
\f{1}{2}{\bf \nabla}^2 \chi = {\cal E} \chi.
\end{eqnarray}
The expansion (1.7),
\begin{eqnarray}
g {\bf S} = g {\bf S}_0 + {\bf S}_1 + g^{-1}{\bf S}_2 + 
g^{-2}{\bf S}_3 + \cdots,
\end{eqnarray}
will now be accompanied by similar expansions for $ \chi $ and ${\cal E}$:
\begin{eqnarray}
\chi = \chi_0+g^{-1}\chi_1+g^{-2}\chi_2+\cdots
\end{eqnarray}
and
\begin{eqnarray}
{\cal E}=g{\cal E}_0+{\cal E}_1+g^{-1}{\cal E}_2+\cdots.
\end{eqnarray}
Substituting (6.5) - (6.7) into (6.4) and equating the coefficients of
 $ g^{-n} $  on both sides, we obtain the following first order partial
 differential equations for $ \chi_0,~\chi_1,~\chi_2,~\cdots $:
\begin{eqnarray}
({\bf \nabla} {\bf S}_0 \cdot {\bf \nabla} - {\cal E}_0) \chi_0 &=&~0,\\
({\bf \nabla} {\bf S}_0 \cdot {\bf \nabla} - {\cal E}_0) \chi_1 &=& 
(- {\bf \nabla} {\bf S}_1 \cdot {\bf \nabla} +
\f{1}{2}{\bf \nabla}^2+{\cal E}_1)\chi_0, \\
({\bf \nabla} {\bf S}_0 \cdot {\bf \nabla} - {\cal E}_0) \chi_2 &=& 
(- {\bf \nabla} {\bf S}_1 \cdot {\bf \nabla} +
\f{1}{2}{\bf \nabla}^2 + {\cal E}_1) \chi_1
+ (-{\bf \nabla} {\bf S}_2 \cdot {\bf \nabla} + {\cal E}_2) \chi_0,
\end{eqnarray}
etc. To see  how these equations can be solved we first give a simple 
example and then address the general solution.

\noindent
\underline{Example}~Take the example
\begin{eqnarray}
v ({\bf q})=\f{1}{2}(\nu_1^2 q_1^2 + \nu_2^2 q_2^2 + \cdots + 
\nu_N^2 q_N^2),
\end{eqnarray}
and correspondingly,
\begin{eqnarray}
{\bf S}_0({\bf q})&=&\f{1}{2}(\nu_1 q_1^2 + \nu_2 q_2^2 + \cdots + 
\nu_N q_N^2)\nonumber\\
{\bf S}_1({\bf q})&=&{\bf S}_2({\bf q})= \cdots = 0,
\end{eqnarray}
\begin{eqnarray}
E=\f{g}{2}(\nu_1 + \nu_2 + \cdots + \nu_N).
\end{eqnarray}
Introduce, as in (1.26)-(1.28), the variables 
$$ ({\bf S}_0,\alpha) $$
with $\alpha=(\alpha_1,\alpha_2,\cdots, \alpha_{N-1})$
and
\begin{eqnarray}
{\bf \nabla} {\bf S}_0 \cdot {\bf \nabla} \alpha_j = 0.
\end{eqnarray}
In terms of these new coordinates $ {\bf S}_0 $ and $ \alpha $, (6.8)
becomes
\begin{eqnarray}
({\bf \nabla} {\bf S}_0)^2(\f{\partial {\rm ln} \chi_0}
{\partial {\bf S}_0})_{\alpha}={\cal E}_0,
\end{eqnarray}
which will be integrated along the constant-$ \alpha $ trajectory. The
constant-$ \alpha $ trajectory is determined by the solution of
the Hamilton-Jacobi equation (1.11); it describes a classical 
trajectory with $-v$ as the potential and a positive infinitesimal 
energy, starting from $ {\bf q}={\bf 0} $ at time near $-\infty$, and 
$ {\bf q}=(q_1,~q_2,~\cdots,~q_N)=({\bf S}_0, \alpha) $ at time $ t $.
As $ t $ increases to $t + dt$, the end of
the trajectory moves from $ {\bf q}=({\bf S}_0, \alpha ) $ to
\begin{eqnarray}
(q_1+dq_1, q_2+dq_2,\cdots,q_N+dq_N)=({\bf S}_0+d{\bf S}_0,\alpha)
\end{eqnarray}
keeping $ \alpha $ constant. In accordance with (2.20) - (2.22), we have
\begin{eqnarray}
dt = \f{d {\bf S}_0}{({\bf \nabla} {\bf S}_0)^2} = 
\f{d q_1}{\nu_1 q_1} = \f{d q_2}{\nu_2q_2} = \cdots =\f{dq_N}{\nu_N q_N}.
\end{eqnarray}
When $ {\bf S}_0 \rightarrow 0 $, each $ q_i $ must also $ \rightarrow 0 $
along the trajectory. Since $ \chi_0 $ is a
single-valued function of ${\bf q} $, we can classify $ \chi_0 $ by its 
power dependence on $ q_i $, as $ q_i\rightarrow 0 $. Take this to be
\begin{eqnarray}
\chi_0 \rightarrow q_1^{n_1}q_2^{n_2}\cdots q_N^{n_N}.
\end{eqnarray}
Since
\begin{eqnarray}
(\f{\partial {\rm ln} \chi_0}{\partial {\bf S}_0})_\alpha = 
\sum\limits^N_{i=1}
(\f{\partial {\rm ln} q_i}{\partial {\bf S}_0})_\alpha
\f{\partial {\rm ln} \chi_0}{\partial {\rm ln} q_i}.
\end{eqnarray}
From (6.17), we see that along the classical (constant $ \alpha $ )
trajectory
\begin{eqnarray}
(\f{\partial {\rm ln} q_i}{\partial {\bf S}_0})_\alpha =
\f{\nu_i}{({\bf \nabla}
{\bf S}_0)^2},
\end{eqnarray}
and consequently
\begin{eqnarray}
(\f{\partial {\rm ln} \chi_0}{\partial {\bf S}_0})_\alpha = 
\f{1}{({\bf \nabla} {\bf S}_0)^2} \sum\limits^N_{i=1}\nu_i
\f{\partial {\rm ln} \chi_0}{\partial {\rm ln} q_i}.
\end{eqnarray}
As $ {\bf S}_0\rightarrow 0 $, under the assumption (6.18), we have
\begin{eqnarray}
({\bf \nabla} {\bf S}_0)^2 
(\f{\partial {\rm ln} \chi_0}{\partial {\bf S}_0})_{\alpha}
\rightarrow \sum\limits^N_{i=1}n_i\nu_i;
\end{eqnarray}
therefore (6.15) yields
\begin{eqnarray}
{\cal E}_0=\sum\limits^N_{i=1} n_i\nu_i
\end{eqnarray}
and
\begin{eqnarray}
({\bf \nabla} {\bf S}_0)^2(\f{\partial {\rm ln} \chi_0}
{\partial {\bf S}_0})_\alpha
= \sum\limits^N_{i=1} n_i \nu_i
\end{eqnarray}
along the entire trajectory for all $ {\bf S}_0 > 0 $. 
(It is reassuring that (6.23) is independent of $\alpha$.)
It follows then
\begin{eqnarray}
\chi_0=q_1^{n_1}q_2^{n_2}\cdots q_N^{n_N}
\end{eqnarray}
at all ${\bf q}$.

In this example, because ${\bf S}_1 = 0$, (6.9) becomes
\begin{eqnarray}
({\bf \nabla} {\bf S}_0)^2
(\f{\partial \chi_1}{\partial {\bf S}_0})_{\alpha}
 - {\cal E}_0 \chi_1 = 
{\cal E}_1 \chi_0 + \f{1}{2}{\bf \nabla}^2\chi_0, \nonumber
\end{eqnarray}
which, on account of (6.15) and (6.25), can also be written as 
\begin{eqnarray}
\chi_0 ({\bf \nabla} {\bf S}_0)^2[\f{\partial}{\partial {\bf S}_0}
(\f{\chi_1}{\chi_0})]_{\alpha} = {\cal E}_1 \chi_0
+\f{1}{2}[\f{n_1(n_1-1)}{q_1^2} +\f{n_2(n_2-1)}{q_2^2} +
\cdots + \f{n_N(n_N-1)}{q_N^2}]\chi_0.
\end{eqnarray}
Thus, keeping $\alpha$ fixed,
\begin{eqnarray}
\chi_1 = \chi_0 \int \f{d {\bf S}_0}
{({\bf \nabla} {\bf S}_0)^2} 
\{{\cal E}_1 + \f{1}{2}
[\f{n_1(n_1-1)}{q_1^2} +\f{n_2(n_2-1)}{q_2^2} +
\cdots + \f{n_N(n_N-1)}{q_N^2}]\},
\end{eqnarray}
where the integration is along the (classical) trajectory normal to the 
${\bf S}_0 =$ constant surfaces. 
We leave (6.27) in the indefinite integral form, since the integration 
constant can be eliminated by the transformation
\begin{eqnarray}
\chi_1 \rightarrow \chi_1 + {\sf constant}\cdot \chi_0
\end{eqnarray}
which affects only the overall normalization factor of $\chi$. (This 
convention will be adopted below for the general case as well.)

For any ${\cal E} \not= 0$, and any partition
\begin{eqnarray}
{\cal E}_1=m_1\nu_1 + m_2\nu_2 + \cdots + m_N\nu_N \nonumber
\end{eqnarray}
(where $m_i$ can be an arbitrary number) gives
\begin{eqnarray}
\int\f{d{\bf S}_0}{({\bf \nabla} {\bf S}_0)^2} {\cal E}_1 = 
m_1 {\rm ln} q_1 + m_2 {\rm ln} q_2 + \cdots + m_N {\rm ln} q_N.\nonumber
\end{eqnarray}
This leads to an inadmissible wave function, and consequently
\begin{eqnarray}
{\cal E}_1 = 0.
\end{eqnarray}
Hence, by using (6.17), (6.27) and (6.29), we find for this example
\begin{eqnarray}
\chi_1 = -\f{\chi_0}{4}
[\f{n_1(n_1-1)}{\nu_1q_1^2} +\f{n_2(n_2-1)}{\nu_2q_2^2} +
\cdots + \f{n_N(n_N-1)}{\nu_Nq_N^2}].
\end{eqnarray}
Recalling that the Hermite polynomial $H_n(x)$ is
\begin{eqnarray}
H_n(z) = (2z)^n [1-\f{n(n-1)}{4z^2} + \cdots],\nonumber
\end{eqnarray}
the sum
\begin{eqnarray}
\chi = \chi_0 + g^{-1}\chi_1 + \cdots\nonumber
\end{eqnarray}
can be shown to be, apart from an overall normalization factor,
\begin{eqnarray}
H_{n_1}(\sqrt{g\nu_1} q_1) H_{n_2}(\sqrt{g\nu_2} q_2) \cdots
H_{n_N}(\sqrt{g\nu_N} q_N).
\end{eqnarray}

Returning to the general case, because ${\bf q} = {\bf 0} $ is the minimum 
of $v({\bf q})$, as ${\bf q} \rightarrow {\bf 0}$, $v({\bf q})$ depends 
quadratically on ${\bf q}$. Write
\begin{eqnarray}
v ({\bf q}) \rightarrow \f{1}{2}(\nu_1^2 q_1^2 + \nu_2^2 q_2^2 + \cdots + 
\nu_N^2 q_N^2);
\end{eqnarray}
correspondingly,
\begin{eqnarray}
{\bf S}_0({\bf q}) \rightarrow \f{1}{2}(\nu_1 q_1^2 + \nu_2 q_2^2 + \cdots + 
\nu_N q_N^2).
\end{eqnarray}
Again, we shall classify $\chi_0({\bf q})$ according to its behavior as 
${\bf q} \rightarrow {\bf 0}$, by using (6.18). As in this example, this 
leads to
\begin{eqnarray}
{\cal E}_0=n_1\nu_1 + n_2\nu_2 + \cdots + n_N\nu_N, 
\end{eqnarray}
where, as before, $n_1,~n_2,~\cdots,~n_N$ are positive integers. From (6.8),
it follows that
\begin{eqnarray}
{\rm ln}\chi_0 = {\cal E}_0 \int \f{d {\bf S}_0}
{({\bf \nabla} {\bf S}_0)^2} 
\end{eqnarray}
with the integration taken along the trajectory of constant $\alpha$, and 
the integration constant is determined by the normalization condition
(6.18), as ${\bf S}_0 \rightarrow 0$.

It is convenient to characterize the limit  ${\bf S}_0 \rightarrow 0$, by
introducing along the trajectory an overall scale factor $\lambda$ (of the
dimension $[q_i]$). Because of (6.17), for sufficiently small ${\bf S}_0$,
we set 
\begin{eqnarray}
{\bf S}_0 \propto \lambda^2,~~~~~~q_i \propto \lambda,
\end{eqnarray}
and therefore, on account of (6.17), 
\begin{eqnarray}
\f{d {\bf S}_0}{({\bf \nabla} {\bf S}_0)^2} = 
\f{1}{\nu_1} d {\rm ln} q_1 = 
\f{1}{\nu_2} d {\rm ln} q_2 = \cdots = 
\f{1}{\nu_N} d {\rm ln} q_N \propto d {\rm ln} \lambda.
\end{eqnarray}
We now turn to (6.9); as in (6.26)-(6.27), it can be written as
\begin{eqnarray}
\chi_0 ({\bf \nabla} {\bf S}_0)^2[\f{\partial}{\partial {\bf S}_0}
(\f{\chi_1}{\chi_0})]_{\alpha} = 
(\f{1}{2}{\bf \nabla}^2 
- {\bf \nabla} {\bf S}_1 \cdot {\bf \nabla})\chi_0 +
{\cal E}_1 \chi_0,\nonumber
\end{eqnarray}
and therefore
\begin{eqnarray}
\chi_1 = \chi_0 \int \f{d {\bf S}_0}
{({\bf \nabla} {\bf S}_0)^2} [
\f{1}{\chi_0}(\f{1}{2}{\bf \nabla}^2 
- {\bf \nabla} {\bf S}_1 \cdot {\bf \nabla}) \chi_0
+ {\cal E}_1].
\end{eqnarray}
Near ${\bf S}_0 = 0$,  we may use (6.36)  to expand the first term inside
the square bracket as a power series in $\lambda$:
\begin{eqnarray}
\f{1}{\chi_0} (\f{1}{2}{\bf \nabla}^2 
- {\bf \nabla} {\bf S}_1 \cdot {\bf \nabla}) \chi_0 &=&
b_{-2} \lambda^{-2} + b_{-1} \lambda^{-1} + b_0 +\nonumber \\
&& b_1 \lambda + b_2 \lambda_2 + \cdots,
\end{eqnarray}
where $b_{-2},~b_{-1},~b_0,~b_1,~\cdots$ are constants. Because of (6.37),
in order that $\chi_1$ be analytic at ${\bf q} = {\bf 0}$, we must require
\begin{eqnarray}
{\cal E}_1 = -b_0;
\end{eqnarray}
otherwise, $\chi_1$ would have a term proportional to  
$\chi_0 {\rm ln} \lambda$, which is not admissible. The function 
$\chi_1$ is given by the integral (6.38) along the $\alpha$-constant 
trajectory. In a similar way, we can determine 
$ {\cal E}_2,~{\cal E}_3,~\cdots$ and obtain the solutions 
$\chi_2,~\chi_3,~\cdots$, in terms of quadrature along the classical 
(constant $\alpha$) trajectory.


\noindent
\section*{\bf 7. Perturbation Around An Attractive Coulomb Potential}
\setcounter{section}{7}
\setcounter{equation}{0}

Let $H_c$ be the Hamiltonian for a Coulomb potential:
\begin{eqnarray}\label{e6.41}
H_c = -\frac{1}{2} {\bf \nabla}^2 - \frac{g^2}{r},
\end{eqnarray}
where 
\begin{eqnarray}\label{e6.42}
g^2 = Z  e^2
\end{eqnarray}
with  ${\bf \nabla}^2 $ denoting the three-dimensional Laplacian, $r$
the radius, $Ze$ the nuclear charge and $-e$ the electronic charge.
Consider a problem in which there is an additional perturbation 
$\epsilon U$; the corresponding Hamiltonian is
\begin{eqnarray}\label{e6.43}
H = H_c + \epsilon U,
\end{eqnarray}
where $U$ is not singular at the origin. Let $\psi_c$ and $\psi$ be
the ground-states of $H_c$ and $H$; i.e.,
\begin{eqnarray}\label{e6.44}
H_c \psi_c = E_c \psi_c
\end{eqnarray}
and
\begin{eqnarray}\label{e6.45}
H \psi = E \psi.
\end{eqnarray}

\noindent
7.1~~Isotropic Case

We first discuss the case that $U(r)$ depends only on $r$.
As we shall see, with modification our method can be adapted to derive 
$\psi$ by quadratures along the radial trajectory.
The solution to the Coulomb problem is well known:
\begin{eqnarray}\label{e6.46}
\psi_c = e^{-g^2r}~~~~~{\sf and} ~~~~~E_c =-\frac{1}{2}g^4.
\end{eqnarray}
This suggests[10] that instead of (1.7)-(1.8), a different 
$g$-power expansion is needed, one that should conform to the form 
(\ref{e6.46}) of the Coulomb wave function.

Write
\begin{eqnarray}\label{e6.47}
\psi = e^{-S}.
\end{eqnarray}
Expand
\begin{eqnarray}\label{e6.48}
S = g^2 S_0 + S_1 + g^{-2} S_2 + g^{-4} S_3 + \cdots + g^{-(2n-2)} S_n
+ \cdots
\end{eqnarray}
and
\begin{eqnarray}\label{e6.49}
E = g^4 E_0 + g^2 E_1 + E_2 + g^{-2} E_3 + \cdots + g^{-(2n-4)} E_n
+ \cdots.
\end{eqnarray}
Since
$$
{\bf \nabla}^2 \psi = [({\bf \nabla}S)^2 -  {\bf \nabla}^2 S] \psi,
$$
(\ref{e6.45}) becomes
\begin{eqnarray}\label{e6.50}
-\frac{1}{2}({\bf \nabla}S)^2 + \frac{1}{2} {\bf \nabla}^2 S 
-\frac{g^2}{r} + \epsilon U =  E.
\end{eqnarray}
Substituting (\ref{e6.48})-(\ref{e6.49}) into (\ref{e6.50}) and equating
the coefficients of $g^4,~g^2,~g^0,~\cdots,~g^{-2m},~\cdots$ on both 
sides, we obtain
\begin{eqnarray}\label{e6.51}
({\bf \nabla} S_0)^2 &=& -2 E_0\\ 
{\bf \nabla} S_0 \cdot {\bf \nabla} S_1 &=& 
\f{1}{2}{\bf \nabla}^2 S_0 - \f{1}{r} - E_1\\
{\bf \nabla} S_0 \cdot {\bf \nabla} S_2 &=& 
-\f{1}{2}({\bf \nabla} S_1)^2 +
\f{1}{2}{\bf \nabla}^2 S_1 + \epsilon U - E_2,\\
{\bf \nabla} S_0 \cdot {\bf \nabla} S_3 &=& 
-{\bf \nabla} S_1 \cdot {\bf \nabla} S_2 +
\f{1}{2}{\bf \nabla}^2 S_2 - E_3,
\end{eqnarray}
etc.; for $n>1$,
\begin{eqnarray}\label{e6.55}
{\bf \nabla} S_0 \cdot {\bf \nabla} S_{2n} =
-\sum\limits_{m=1}^{n-1}{\bf \nabla} S_m \cdot {\bf \nabla} S_{2n-m} 
-\f{1}{2} ({\bf \nabla} S_n)^2 + \f{1}{2} {\bf \nabla}^2 S_{2n-1} - E_{2n}
\end{eqnarray}
and for $n \geq 1$
\begin{eqnarray}\label{e6.56}
{\bf \nabla} S_0 \cdot {\bf \nabla} S_{2n+1} =
-\sum\limits_{m=1}^n{\bf \nabla} S_m \cdot {\bf \nabla} S_{2n+1-m} 
+ \f{1}{2} {\bf \nabla}^2 S_{2n} - E_{2n+1}.
\end{eqnarray}

For the ground-state, these $S_m(r) $ are all radial functions. Since
$U(r)$ is regular at $r=0$, we can set
\begin{eqnarray}\label{e6.57}
U(0) = 0.
\end{eqnarray}
We adopt the  same normalization  condition (1.36), with
\begin{eqnarray}\label{e6.58}
\psi(0) = 1~~~~~{\sf and}~~~~~~S(0) = 0.
\end{eqnarray}
From  (\ref{e6.51}), it follows that
\begin{eqnarray}\label{e6.59}
&\frac{\partial S_0}{\partial r} = \sqrt{-2 E_0}\nonumber\\
{\sf and}~~~~~~~~~~~~~~&\\
&S_0 = \sqrt{-2 E_0}~r.\nonumber
\end{eqnarray}
Substituting these expressions into (7.12), we  find
\begin{eqnarray}\label{e6.60}
\sqrt{-2 E_0} \frac{\partial S_1}{\partial r} = 
\frac{1}{r} (\sqrt{-2 E_0} - 1) - E_1.
\end{eqnarray}
Because $S_1$ should be regular at $r =  0$, $\sqrt{-2 E_0} - 1 = 0$
and therefore[10]
\begin{eqnarray}\label{e6.61}
E_0 = -\frac{1}{2}~~~~~{\sf and}~~~~~~S_0 = r
\end{eqnarray}
confirming (\ref{e6.46}); (\ref{e6.60}) then leads to
\begin{eqnarray}\label{e6.62}
&\frac{\partial S_1}{\partial r} = - E_1 \nonumber\\  
{\sf and}~~~~~~~~~~~~~~&\\
&S_1 = - E_1~r.\nonumber
\end{eqnarray}
Since
\begin{eqnarray}\label{e6.63}
{\bf \nabla}^2 r=\frac{2}{r},
\end{eqnarray}
we see  that (7.13) becomes
\begin{eqnarray}
\frac{\partial S_2}{\partial r} = - \frac{1}{2}E_1^2  - 
\frac{E_1}{r} + \epsilon U(r) - E_2. \nonumber
\end{eqnarray}
In order that  $S_2(r)$ be regular at $r=0$,
\begin{eqnarray}\label{e6.64}
E_1  = 0,
\end{eqnarray}
and consequently, on account of (\ref{e6.62}),
\begin{eqnarray}\label{e6.65}
S_1(r)  = 0
\end{eqnarray}
and
\begin{eqnarray}\label{e6.66}
\frac{\partial S_2}{\partial r} = \epsilon U(r) - E_2. 
\end{eqnarray}
By using (\ref{e6.63}), we see that as $r \rightarrow 0$
\begin{eqnarray}\label{e6.67}
{\bf \nabla}^2 S_2 \rightarrow ({\bf \nabla}^2 S_2)_0 =
\frac{2}{r} (\frac{\partial S_2}{\partial r})_0 = 
\frac{2}{r} [\epsilon U(0) - E_2]. 
\end{eqnarray}
From (7.14), we see that in order to have $S_3$ also regular at
$r=0$, $(\frac{\partial S_2}{\partial r})_0$ must be $0$; this yields
on account of (\ref{e6.57}),
\begin{eqnarray}\label{e6.68}
E_2  &=& \epsilon U(0) = 0\\
\frac{\partial S_2}{\partial r} &=& \epsilon U(r) 
\end{eqnarray}
and
\begin{eqnarray}\label{e6.70}
\frac{\partial S_3}{\partial r} =
\frac{\epsilon}{2r^2} \frac{\partial}{\partial r} (r^2 U) - E_3.
\end{eqnarray}
As $r \rightarrow 0$, (7.14) implies
\begin{eqnarray}\label{e6.71}
\frac{\partial S_3}{\partial r} \rightarrow
(\frac{\partial S_3}{\partial r})_0 =
[\frac{\epsilon}{2r^2} \frac{\partial}{\partial r} (r^2 U)]_0 - E_3
\end{eqnarray}
and, on account of (\ref{e6.63}),
\begin{eqnarray}\label{e6.72}
{\bf \nabla}^2 S_3 \rightarrow
\frac{2}{r} (\frac{\partial S_3}{\partial r})_0.
\end{eqnarray}
Hence by using (\ref{e6.55}) for $S_4$ and in order that $S_4$ should be
regular at $r=0$,
\begin{eqnarray}\label{e6.73}
(\frac{\partial S_3}{\partial r})_0 = 0,
\end{eqnarray}
and therefore
\begin{eqnarray}\label{e6.74}
E_3 = \frac{\epsilon}{2}[\frac{1}{r^2}
\frac{\partial}{\partial r}(r^2 U)]_0. 
\end{eqnarray}
The same reasoning yields for all $m \geq 1$, at $r=0$,
\begin{eqnarray}\label{e6.75}
(\frac{\partial S_m}{\partial r})_0 = 0.
\end{eqnarray}
From (\ref{e6.74}), we see that for
\begin{eqnarray}\label{e6.76}
\epsilon U(r) &=& \epsilon r^l\nonumber \\
E_3 &=& 
\left\{
\begin{array}{ll}
\frac{3}{2} \epsilon~~~~~~&{\sf if}~~~~~~l=1 \\
0                         &{\sf if}~~~~~~l>1.
\end{array}
\right.
\end{eqnarray}

\noindent
\underline{Example}

To  carry the analysis further, let us consider the case $l=2$; i.e.,
\begin{eqnarray}\label{e6.77}
\epsilon U(r) = \epsilon r^2.
\end{eqnarray}
In this example, the above results give
\begin{eqnarray}\label{e6.78}
E_1 = E_2 = E_3 = 0\nonumber\\
S_1 =0,~~~~~~~S_2 = \frac{1}{3} \epsilon r^3 
\end{eqnarray}
and
$$
S_3 = \epsilon r^2.
$$
Set $n=2$ in (\ref{e6.55}), we have
\begin{eqnarray}
\frac{\partial S_4}{\partial r} = 
-\frac{1}{2}\epsilon^2 r^4 + 3\epsilon - E_4.\nonumber
\end{eqnarray}
Because of (\ref{e6.75}), at $r=0$, $\frac{\partial S_4}{\partial r} = 0$; 
hence,
\begin{eqnarray}\label{e6.79}
E_4 = 3\epsilon 
\end{eqnarray}
and
\begin{eqnarray}\label{e6.80}
S_4 = -\frac{1}{10} \epsilon^2 r^5. 
\end{eqnarray}
In  a similar way, we derive
\begin{eqnarray}\label{e6.81}
E_5 = 0,~~~~~~~~&S_5 = -\frac{7}{8} \epsilon^2 r^4,\nonumber\\
E_6 = 0,~~~~~~~~&S_6 = -\frac{43}{12} \epsilon^2 r^3 +
\frac{1}{14} \epsilon^3 r^7,\nonumber\\
E_7 = 0,~~~~~~~~&S_7 = -\frac{43}{4} \epsilon^2 r^2 +
\frac{13}{12} \epsilon^3 r^6,\\
E_8 = -\frac{129}{4} \epsilon^2,&{\sf etc.}\nonumber
\end{eqnarray}
Putting  together, for $\epsilon U =  \epsilon r^2$, we find
\begin{eqnarray}\label{e6.82}
S &=& g^2r + \frac{\epsilon}{3g^2} r^3 + \frac{\epsilon}{g^4} r^2
-\frac{\epsilon^2}{10g^6} r^5\nonumber\\
&&-\frac{7\epsilon^2}{8g^8} r^4 + \frac{1}{g^{10}} 
(-\frac{43}{12} \epsilon^2 r^3 +\frac{1}{14} \epsilon^3 r^7)\nonumber\\
&&+ \frac{1}{g^{12}} (-\frac{43}{4} \epsilon^2 r^2 +
\frac{13}{12} \epsilon^3 r^6) + O(\frac{\epsilon^3}{g^{14}})
\end{eqnarray}
and
\begin{eqnarray}\label{e6.83}
E = -\frac{g^4}{2} + \frac{3\epsilon}{g^4}
-\frac{129}{4g^{12}} \epsilon^2 +O(\frac{\epsilon^3}{g^{20}}).
\end{eqnarray}
In the usual perturbation series, to each order of the perturbation,
the derivation of the perturbed Coulomb wave function requires summations
over an infinite number of excited bound-states, plus the continuum.
Here, to each order $\epsilon^m$, the perturbed wave function  can be
obtained in  closed form by quadratures along the radial trajectory.

The perturbed energy can also be derived in an alternative way by using
the  integral form (4.92):
\begin{eqnarray}\label{e6.84}
E = -\frac{1}{2}g^4 + \frac
{\int\limits_0^{\infty} e^{-g^2r-S}  \epsilon U r^2 dr}
{\int\limits_0^{\infty} e^{-g^2r-S} r^2 dr}.
\end{eqnarray}
To first order in $\epsilon$, we can approximate the factor $e^{-S}$
by $e^{-g^2r}$ in the integrals. For $U=r^2$,
\begin{eqnarray}\label{e6.85}
E &=& -\frac{1}{2}g^4 + \frac
{\epsilon \int\limits_0^{\infty} e^{-2g^2r} r^4 dr}
{\int\limits_0^{\infty} e^{-2g^2r} r^2 dr} +O(\epsilon^2)\nonumber\\
&=&  -\frac{1}{2}g^4 + \frac{3\epsilon}{g^4}+O(\epsilon^2).  
\end{eqnarray}
To calculate $E$ to the accuracy of $O(\epsilon^2)$, we need the wave
function to  $O(\epsilon)$. From (\ref{e6.82}),
\begin{eqnarray}\label{e6.86}
e^{-S} =e^{- g^2r}[1 - \epsilon(\frac{r^3}{3g^2} + \frac{r^2}{g^4})
+ O(\epsilon^2)].
\end{eqnarray}
Substituting this expression into (\ref{e6.84}), we find the second
term on its right-hand side to be:
\begin{eqnarray}\label{e6.87}
\frac
{\epsilon \int\limits_0^{\infty} e^{-g^2r}r^4
[1-\epsilon (\frac{r^3}{3g^2} + \frac{r^2}{g^4})] dr}
{\int\limits_0^{\infty} e^{-g^2r}r^2
[1-\epsilon (\frac{r^3}{3g^2} + \frac{r^2}{g^4})] dr},
\end{eqnarray}
confirming (\ref{e6.83}).

\noindent
7.2~~Stark Effect

Consider the anisotropic example, in which the perturbation is
\begin{eqnarray}\label{e7.48}
\epsilon U = \epsilon r {\rm cos}a,
\end{eqnarray}
where $a$ is the polar angle; i.e., 
\begin{eqnarray}\label{e7.49}
r^2 = x^2 + y^2 + z^2~~~~{\sf and}~~~~~z = r {\rm cos}a.
\end{eqnarray}
Replacing $\epsilon U(r)$ by $ \epsilon r {\rm cos}a$, we see that 
(\ref{e6.41})-(\ref{e6.68}) remain intact with $E_1=E_2=S_1=0$;
(7.29) becomes
\begin{eqnarray}\label{e7.50}
\frac{\partial S_2}{\partial r} = \epsilon~r~{\rm cos}a,
\end{eqnarray}
and therefore
\begin{eqnarray}\label{e7.51}
S_2 = \frac{1}{2} \epsilon~r^2 {\rm cos}a,
\end{eqnarray}
and
\begin{eqnarray}\label{e7.52}
{\bf \nabla}^2 S_2 = 2~\epsilon~{\rm cos}a.
\end{eqnarray}
Thus, (7.14) gives
\begin{eqnarray}\label{e7.53}
\frac{\partial S_3}{\partial r} = \epsilon~{\rm cos}a - E_3,
\end{eqnarray}
which yields, on account of (7.49),
\begin{eqnarray}\label{e7.54}
S_3 = \epsilon~z - E_3 r
\end{eqnarray}
and
\begin{eqnarray}\label{e7.55}
{\bf \nabla}^2 S_3 = - \frac{2 E_3}{r}.
\end{eqnarray}
Since $S_4$ satisfies (7.15) for $n=2$,
\begin{eqnarray}
\frac{\partial S_4}{\partial r} = -\frac{1}{2} ({\bf \nabla} S_2)^2 + 
\frac{1}{2} {\bf \nabla}^2 S_3 - E_4.\nonumber
\end{eqnarray}
In order that $S_4$ not have a ${\rm ln}r$ singularity,
\begin{eqnarray}
E_3 = 0\nonumber
\end{eqnarray}
and therefore
\begin{eqnarray}\label{e7.56}
S_3 = \epsilon~z.
\end{eqnarray}
In order that ${\bf \nabla}^2 S_4$ not have a $1/r$ singularity,
\begin{eqnarray}\label{e7.57}
E_4 = 0.
\end{eqnarray}
Thus,
\begin{eqnarray}\label{e7.58}
S_4 = -\frac{1}{24} \epsilon^2 r^3 (1 + 3~{\rm cos}^2 a).
\end{eqnarray}
Likewise, it is straightforward to derive
\begin{eqnarray}\label{e7.59}
E_5 &=& 0,~~~~~~S_5 = -\frac{7}{16} \epsilon^2 r^2 (1 + {\rm cos}^2 a),\\
E_6 &=& -\frac{9}{4}\epsilon^2,~~~S_6 = \frac{1}{16} \epsilon^3 r^4 
{\rm cos}a (1 + {\rm cos}^2 a),\\
E_7 &=& 0,~~~~~~S_7 = \frac{13}{48} \epsilon^3 r^3 
{\rm cos}a (3 + {\rm cos}^2 a),\\
E_8 &=& 0,~~~~~~S_8 = \frac{53}{16} \epsilon^3 r^2 {\rm cos}a -
\frac{1}{128} \epsilon^4 r^5 (1 + 10 {\rm cos}^2 a + 5 {\rm cos}^4 a),\\
E_9 &=& 0,~~~~~~S_9 = \frac{53}{8} \epsilon^3 r {\rm cos}a -
\frac{99}{512} \epsilon^4 r^4 (1 + 6 {\rm cos}^2 a + {\rm cos}^4 a),\\
E_{10} &=& 0,~~~~~~S_{10} = -\frac{761}{384} \epsilon^4 r^3 
(1+ 3 {\rm cos}^2a) +O(\epsilon^5),\\
E_{11} &=& 0,~~~~~~S_{11} = -\frac{3131}{256} \epsilon^4 r^2
(1+ {\rm cos}^2a) +O(\epsilon^5),\\
E_{12} &=& -\frac{3555}{64} \epsilon^4, 
\end{eqnarray}
etc.

Combining these results together, for the potential
$-\frac{g^2}{r} + \epsilon r {\rm cos}a$, 
we find that, in powers of $\epsilon$, the wave function $e^{-S}$ and the
energy are given by
\begin{eqnarray}\label{e7.67}
S &=& g^2 r + \frac{\epsilon r}{g^4} {\rm cos}a (1 + \frac{1}{2} g^2 r)
- \frac{\epsilon^2 r^2}{g^8} [\frac{7}{16}(1+ {\rm cos}^2a)
+\frac{1}{24} g^2 r (1 + 3 {\rm cos}^2 a)]\nonumber\\
&+& \frac{\epsilon^3 r}{g^{16}} {\rm cos}a
[\frac{53}{8} (1 + \frac{1}{2} g^2 r) +
\frac{13}{48} (g^2r)^2 (3 + {\rm cos}^2 a) +
\frac{1}{16} (g^2r)^3 (1 + {\rm cos}^2 a)]\nonumber\\
&+& O(\epsilon^4)
\end{eqnarray}
and
\begin{eqnarray}\label{e7.68}
E = -\frac{1}{2} g^4 - \frac{9}{4} \frac{\epsilon^2}{g^8}
- \frac{3555}{64}\frac{\epsilon^4}{g^{20}} + O(\epsilon^6).
\end{eqnarray}
The $- \frac{9}{4} \frac{\epsilon^2}{g^8}$ is known[11,12]. Our method 
gives closed expressions for both the wave function and the energy to any
finite order of $\epsilon$.

\newpage

\section*{\bf Acknowledgment}
\setcounter{section}{10}
\setcounter{equation}{0}
One of us (T. D. Lee) wishes to thank T. K. Lee for discussions, and for 
a private communication which outlines how our method can be extended to
include the hydrogen atom. 
We also wish to give our thanks to RIKEN, 
Brookhaven National Laboratory and to the U.S. Department of Energy 
[DE-AC02-98CH10886] for providing the facilities essential for the 
completion of this work.

}
\end{document}